\documentclass[a4paper,11pt]{article}

\usepackage[top=2.5cm, bottom=2.5cm, left=2.5cm, right=2.5cm]{geometry}
\usepackage[sc]{mathpazo}
\usepackage{amsmath,amssymb,mathtools,amsthm}

\usepackage{authblk}

\usepackage{graphicx}
\usepackage[labelsep=endash, figurename=Figure, tablename=Table, textfont=it]{caption}
\usepackage{subcaption}
\usepackage{float}
\usepackage{booktabs}
\usepackage{longtable}

\usepackage[linesnumbered,ruled,vlined]{algorithm2e}

\usepackage[round,comma,authoryear]{natbib}

\usepackage[dvipsnames]{xcolor}

\usepackage{bbm}

\usepackage{hyperref}
\definecolor{darkblue}{rgb}{0,0,0.5}
\hypersetup{
  colorlinks=true,
  linkcolor=black,
  citecolor=darkblue,
  urlcolor=darkblue
}

\DeclareMathOperator*{\E}{\mathcal{E}}

\DeclareMathOperator*{\G}{\mathcal{G}}

\DeclareMathOperator*{\A}{\mathcal{A}}
\DeclareMathOperator*{\N}{\mathcal{N}}
\DeclareMathOperator*{\M}{\mathcal{M}}

\DeclareMathOperator{\interior}{int}

\newtheorem{proposition}{Proposition}
\newtheorem{lemma}{Lemma}

\newtheorem{assumption}{Assumption}

\title{Mobility-as-a-service (MaaS) system as a multi-leader-multi-follower game: \\
A single-level variational inequality (VI) formulation}

\author[a]{Rui Yao\thanks{These authors contributed equally.}}
\author[b]{Xinyu Ma*}
\author[b]{Kenan Zhang\thanks{Corresponding author: \texttt{kenan.zhang@epfl.ch}}}

\affil[a]{Technion - Israel Institute of Technology, Haifa, Israel}
\affil[b]{École Polytechnique Fédérale de Lausanne (EPFL), CH-1015 Lausanne, Switzerland}

\date{}

\makeatletter
\providecommand{\@LN}[2]{}
\providecommand{\@LN@col}[1]{}
\providecommand{\@LN@lab}[1]{}
\makeatother

\begin{document}
\maketitle

\begin{abstract}
This study models a Mobility-as-a-Service (MaaS) system as a multi-leader-multi-follower game that captures the complex interactions among the MaaS platform, service operators, and travelers. 
We consider a coopetitive setting where the MaaS platform purchases service capacity from service operators and sells multi-modal trips to travelers following an origin-destination-based pricing scheme; meanwhile, service operators use their remaining capacities to serve single-modal trips. 
As followers, travelers make both mode choices, including whether to use MaaS, and route choices in the multi-modal transportation network, subject to prices and congestion.
Inspired by the dual formulation for traffic assignment problems, we propose a novel single-level variational inequality (VI) formulation by introducing a \textit{virtual} traffic operator, along with the MaaS platform and multiple service operators. 
A key advantage of the proposed VI formulation is that it supports parallel solution procedures and thus enables large-scale applications.
We prove that an equilibrium solution always exists given the negotiated wholesale price of service capacity. 
Numerical experiments on a small network further demonstrate that the wholesale price can be tailored to align with varying system-wide objectives. The proposed MaaS system demonstrates potential for creating a ``win-win-win'' outcome---service operators and travelers are better off compared to the ``without MaaS'' scenario, meanwhile the MaaS platform remains profitable. Such a Pareto-improving regime can be explicitly specified with the wholesale capacity price. 
Similar conclusions are drawn from the experiment of an extended multi-modal Sioux Falls network, which also validates the scalability of the proposed model and solution algorithm.
\end{abstract}

\noindent\textbf{Keywords:} Mobility-as-a-service; Coopetition; Multi-leader-multi-follower game; Variational Nash equilibrium; Perturbed utility theory.

\section{Introduction}
Mobility-as-a-Service (MaaS) aims to provide travelers with convenient, seamless mobility options by integrating various transportation services~\citep{MARKARD2012955, hensher2023mobility}. With MaaS, travelers can consume mobility as a unified service rather than purchasing individual travel modes separately~\citep{jittrapirom2017mobility}.
Multiple trials have been implemented in various countries, such as the pilots in Australia~\citep{hensher2020sydney} and UK~\citep{longman2022developing}, Whim in Finland, and UbiGo in Sweden~\citep{karlsson2017deliverable}, and received positive feedback from travelers~\citep{smith2018mobility,kamargianni2016critical}. 
The long-term implementation of MaaS, however, is rather challenging because it requires the MaaS platform to integrate service capacities from multiple service operators, often including those in direct competition for travelers. 
This imposes a great complexity in the service design and operations and accordingly, introduces a series of new research problems. 
Some primary questions include: 
i) how to model the interactions among the MaaS platform, service operators, and travelers, ii) how to price MaaS trips in anticipation of travelers' mode and route choice, iii) how service operators would respond to the MaaS platform's operational strategies, iv) how travelers would choose among different alternatives, v) how the revenue split between the MaaS platform and service operators would affect the multi-modal mobility system, and vi) whether the introduction of MaaS can simultaneously benefit all stakeholders meanwhile ensure economic viability.   

Motivated by the above questions, we study the optimal pricing problem of the MaaS platform and multiple service operators that compete for traveler demand in the same multi-modal transportation network. 
Specifically, the MaaS platform is modeled as an intermediary that purchases capacity from service operators and sells multi-modal trips to travelers. 
Meanwhile, service operators use their remaining capacities to serve single-modal trips on their own pricing schemes, competing with MaaS and other operators for travelers. 
Consequently, the MaaS platform and service operators both cooperate and compete with each other, also known as \textit{coopetition}~\citep{brandenburger1996co}. 
Although several recent studies examine the MaaS platform pricing problem \citep[e.g.,][]{liu2024demand,YAO2024102991,xiSingleleaderMultifollowerGames2024a}, all of them assume exogenous pricing for other mobility services. 
Besides, service operators are often assumed to be fully compliant~\citep{ding2023mechanism}, or with limited negotiation power~\citep{YAO2024102991,xiSingleleaderMultifollowerGames2024a,liu2024demand,pantelidis2020many}.
This simplification is largely due to the complexity of modeling the decision-making processes of different stakeholders and their interactions in the MaaS system. 
However, such simplification could lead to overoptimistic predictions of MaaS. 
Although MaaS showcases promising efficiency in the literature~\citep[e.g.,][]{YAO2024102991,banerjee2025plan}, it is hardly economically sustainable in real practice. 
For example, Whim ceased its operations and filed for bankruptcy in 2024. Travel demand in the above-mentioned pilots also quickly dropped after the strong governmental incentives terminated.
The significant gap between research and real-world implementation highlights the need for a comprehensive MaaS model that captures strategic interactions among all stakeholders, enabling more realistic predictions and actionable insights.

To this end, we formulate a multi-leader-multi-follower game, where the MaaS platform and service operators, as leaders, optimize their operational strategies and the travelers, as followers, make mode and route choices in the multi-modal transportation network. 
The follower's problem can also be seen as a multi-modal traffic assignment, where ``congestion'' emerges both physically on the road network and virtually in the access to particular mobility services. 
In this study, we apply the perturbed utility Markovian choice model (PUMCM) to characterize travelers' mode and route choices, which has shown particular advantages in large-scale and complex networks~\citep{yaoperturbed,yao2025integrated}. 
Moreover, we draw inspiration from the dual formulation of traffic assignment problems~\citep{baillon2008markovian} and create a ``virtual'' traffic operator that optimizes link travel time in parallel with the MaaS platform and service operators that optimize trip fares. 
{This ``virtual'' traffic operator aims to match network link \textit{supply} with travel demand by deciding the link travel times as an analogy of the pricing decisions made by other service operators. Specifically, the travel time of congestible links (e.g., road network) will increase with travel demand to ensure supply-demand balance, while for capacitated links with fixed travel time (e.g., transit network), the congestion effect is reflected by the difference between the ``designed'' travel time and ``real'' travel time. Consequently, the optimal decision of the ``virtual'' traffic operator corresponds to the equilibrium link travel times and dual variables.}


The proposed reformulation has several advantages. 
First and foremost, it largely simplifies the solution procedure. 
A classic bi-level formulation of a multi-leader-multi-follower game would characterize the traffic equilibrium at the lower level and formulate the coopetition problem at the upper level~\citep[e.g.,][]{xi2024strategizing,asadabadi2018co}. 
To solve such a bi-level problem~\citep{dempe2020bilevel}, one needs to compute the traffic equilibrium given the upper-level decisions, and then update the upper-level decisions with the lower-level equilibrium and sensitivities~\citep[e.g.,][]{zhang2018mitigating,zhang2021inter}. 
{The evaluations of equilibrium and sensitivities are both computationally expensive---the former typically involves a number of iterations, while the latter is often challenged by the issue of differentiability~\citep{patriksson2004sensitivity}.
Although some stochastic user equilibrium (SUE) models yield a differentiable equilibrium mapping~\citep[e.g.,][]{zhou2005generalized, liu2016modeling}, the resulting bi-level formulation still requires a double-loop solution procedure that is hardly feasible for large-scale problems.}
{The modeling framework proposed in this study, however, mitigates this issue through a different problem formulation. By introducing the ``virtual'' traffic operator and having it update link travel times along with other service operators, we are able to decompose the traffic equilibrium and simplify the lower level as a variant of \textit{stochastic shortest path problem}~\citep{bertsekas2012dynamic}, where travelers make the \textit{perturbed best response}~\citep{hofbauer2002global} towards travel time and pricing decisions given by the upper level.}
Under PUMCM, the travelers' best response corresponds to the optimal routing strategy in the multi-modal network and possesses an explicit and differentiable expression. Furthermore, followers at the lower level, as well as leaders at the upper level, can update their decisions independently, which enables parallel computation in large-scale networks.

In addition to solution simplicity, the proposed formulation leads to a connection between the equilibrium condition and a variational inequality (VI) problem, based on which we establish the existence of equilibrium. Specifically, we prove the MaaS system always possesses a variational Nash equilibrium~\citep{facchinei2007generalized} subject to any viable wholesale capacity prices. 
From a practical perspective, the proposed model also better reflects the current practice of mobility services, where mobility service operators become more flexible with their operations and can dynamically adjust their pricing schemes in response to travel demand and traffic conditions.

In sum, this work contributes to the literature in the following ways:
\begin{itemize}
    \item It models both the cooperation and competition, hence the ``coopetition'', between the MaaS platform and the service operators that both make strategic decisions to compete for travelers in the same multi-modal network.
    \item It casts the multi-leader–multi-follower game into a single-level variational inequality problem by introducing a ``virtual'' traffic operator as an additional leader. Consequently, it only requires the followers to make the best responses towards travel times and pricing decisions given by the leaders, circumventing the need to compute lower-level equilibrium and to evaluate equilibrium sensitivities.
    \item It shows that, for any viable wholesale capacity price and under mild conditions, there exists a variational Nash equilibrium of the multi-leader–multi-follower game in the MaaS system.
    \item Numerical experiments demonstrate, with a proper capacity wholesale price, the potential of MaaS to achieve a ``win–win–win'' outcome, where the MaaS platform remains profitable, and both service operators and travelers are better off compared with the scenario without MaaS.
\end{itemize}

The remainder of this paper is organized as follows. Section~\ref{sec:network} defines the multi-modal network used in this study. Section~\ref{sec:game} introduces the multi-leader–multi-follower game and formulates the problem for each leader and follower. 
Section~\ref{sec:equilibrium} presents the equilibrium conditions, establishes its existence, and describes the solution algorithm implemented in the numerical experiments. 
Section~\ref{sec:experiment} reports the numerical experiments on a small network with in-depth analyses and on a large network to verify the model and solution scalability. Finally, Section~\ref{sec:conclusion} summarizes the main findings and concludes this study with future directions. 

\section{Multi-modal network}\label{sec:network}

Let $\M$ denote the set of service operators. We consider each operator $m \in \M$ serves a MaaS subnetwork, denoted as $G_m$, and a non-MaaS subnetwork, denoted as $\tilde{G}_m$, with the same fleet (``$\sim$'' is used to denote non-MaaS variables hereafter). These subnetworks collectively build the MaaS and non-MaaS subnetworks denoted as ${G} \coloneqq ({\N}, {\E})$, and $\tilde{G} \coloneqq (\tilde{\N}, \tilde{\E})$, respectively. 
The operator-specific subnetworks are connected with access links within the MaaS and non-MaaS subnetworks. However, the MaaS and non-MaaS subnetworks are mutually exclusive.
Besides, we define a driving subnetwork and denote it as $\check{G} \coloneqq (\check{\N}, \check{\E})$.

\begin{figure}[htb]
  \centering
  \includegraphics[width=0.85\textwidth]{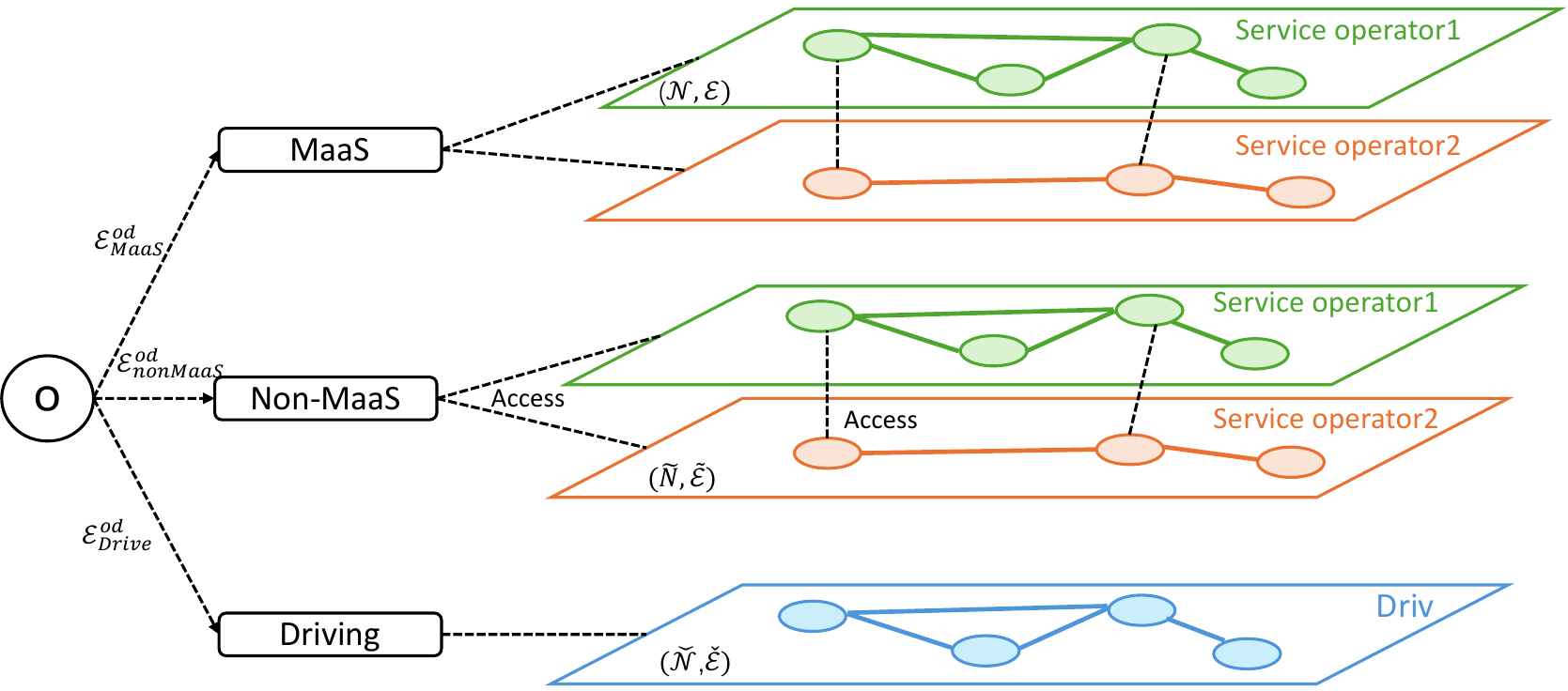} 
  \caption{Multi-modal network \\(the operators serve both MaaS and non-MaaS travelers with the same fleet, and travelers face different fares in MaaS and non-MaaS)}
  \label{fig:MaaS network}
\end{figure}

The MaaS, non-MaaS, and driving subnetworks build up the main body of the multi-modal network denoted as $\G\coloneqq (\mathcal{V}, \mathcal{L})$. 
To connect them with trip origins and destinations, we further define a set of origin nodes, denoted by $\N_O$. Each origin node is further connected to a set of dummy OD-mode nodes $\N_{OD, mode}$ via a set of dummy links $\E_{\text{dummy}}= \{\E^{od}_\text{MaaS},\E^{od}_\text{nonMaaS}, \E^{od}_\text{Drive}\}_{(o,d) \in \N_O \times \N_D}$ to represent mode choice between MaaS, non-MaaS, and driving for each OD. The dummy OD-mode nodes are then connected with service nodes in the subnetworks with access links. 
{In the current setup, there is no transfer link between MaaS, non-MaaS, and driving subnetworks, implying that travelers cannot switch between them during the trips, but can combine different services (e.g., taxi, bus) within each subnetwork.}
An example of the multi-modal network is illustrated in Figure~\ref{fig:MaaS network}.



{The multi-modal transportation network consists of multiple duplicates of the operator's physical service network. Specifically, the service network of operator $m$ is copied for MaaS and non-MaaS, and it is further duplicated for all OD pairs to model different pricing schemes. Besides, we assume that driving (i.e., traveling with private cars) is available to all travelers as an opt-out option, and thus it also leads to multiple copies of road networks. As a result, the size of the multi-modal transportation network expands rapidly with the size of physical networks and the number of OD pairs. For example, the Sioux Falls network used in our experiments consists of 76 base links, but yields 11,018 links when expanded with multi-modal travel options.
Furthermore, strong link interactions occur in the multi-modal network. For instance, MaaS and non-MaaS travelers jointly determine the link flows in the physical networks. Additionally, the MaaS and non-MaaS travel flows of on-demand services and the driving flows together yield the total link flows on the road network. 
Such a significant growth in network size with complex link interactions thus highlights the need for a scalable model and algorithm. 

}

\section{MaaS system as a multi-leader-multi-follower game}\label{sec:game}

In this study, we consider a MaaS platform as an \emph{intermediary} in the multi-modal mobility system that purchases service capacity from multiple service operators and sells multi-modal trips to travelers~\citep{van2022business}. Following \cite{YAO2024102991}, we assume the MaaS platform implements an origin-destination-based pricing. It thus provides travelers with maximum flexibility in choosing any modes and routes to complete their trips.
Differently, we do not model the decision-making process of MaaS trip fare and the wholesale price of service capacity at the same stage. Instead, we assume the MaaS platform and service operators first negotiate the wholesale price per unit utilized capacity by service type, such that the total payment made by the MaaS platform to each service operator can be easily computed based on the realized travel flows. 
In what follows, we focus on the pricing problems at the operational level while leaving the negotiation of wholesale capacity price to future research. Nevertheless, we perform a comprehensive sensitivity analysis on the wholesale capacity price in Section~\ref{sec:experiment-small-capacity} to investigate its impact on the MaaS system.

\begin{figure}[htb]
    \centering
    \includegraphics[width=0.8\textwidth]{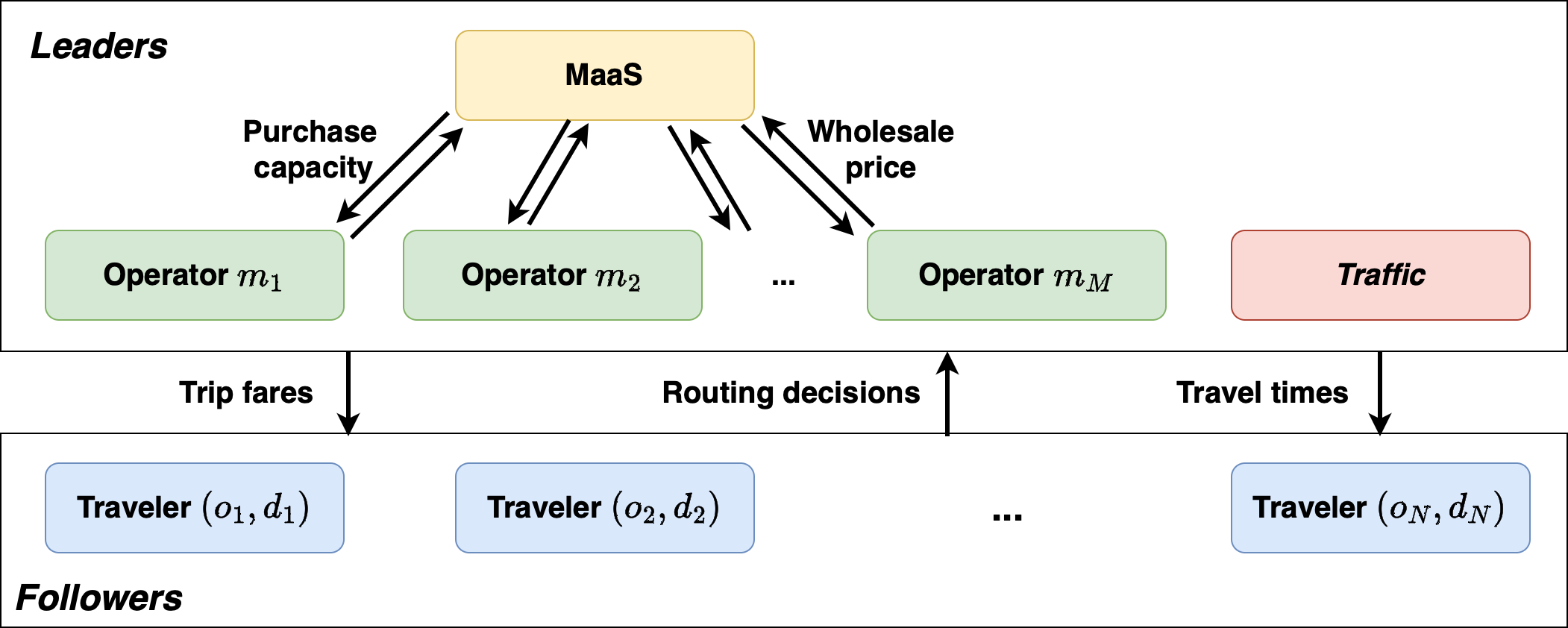}
    \caption{Multi-leader-multi-follower game for a MaaS system.}
    \label{fig:MaaS_multi_leaders}
\end{figure}

The MaaS system envisioned in this study is illustrated in Figure~\ref{fig:MaaS_multi_leaders}. The main stakeholders are classified into two groups: i) Followers: include all travelers in the multi-modal network making both mode choices and route choices to maximize expected travel utilities; and ii) Leaders: consist of all service operators, the MaaS platform, and a \emph{virtual} traffic operator. 
Specifically, travelers can freely choose between driving, non-MaaS, and MaaS, while the total travel demand is fixed and exogenous. 
Both non-MaaS and MaaS travelers take self-designed single- or multi-modal trips, while differing in the consequent trip fares and transfer costs. The routing problem of travelers will be detailed in Section~\ref{sec:follower}. 
The MaaS platform and service operators are all profit-driven and optimize their pricing strategies in anticipation of traveler behaviors and the other operators' response, while the traffic operator ``handles'' the congestion effects and capacity constraints in the multi-modal network. 
Their corresponding problems will be specified in Sections~\ref{sec:traffic operator}-\ref{sec: mobility operator}, respectively. 
Throughout this paper, we assume the service capacities are fixed and exogenous, and the multi-modal network in total has sufficient capacity to sustain all travel demand. 
Collectively, the MaaS system can be formulated as a multi-leader-multi-follower game, and its equilibrium conditions will be formally established in Section~\ref{sec:equilibrium}. 

\subsection{Followers: Travelers} \label{sec:follower}

Following the perturbed utility Markov choice model (PUMCM)~\citep{yaoperturbed}, we model each traveler's routing decisions in the multi-modal network as an infinite-horizon Markov decision process (MDP) with a termination state $d\in \N_D$ (i.e., destination node). {PUMCM is a generalization of classic recursive logit models~\citep{fosgerau2013link,mai2015nested,oyama2022markovian} that allows for more flexible behavioral patterns and supports more efficient solution algorithms.  
Specifically, under PUMCM, the link choice probabilities can be obtained in closed form after simple value iteration that guarantees global convergence, overcoming key challenge in existing recursive models~\citep{mai2022undiscounted}. 
Moreover, the link flows and their sensitivities towards link utilities can also be obtained from the link choice probabilities in closed-form. 
These properties thus make PUMCM an ideal foundation for any upper-level design problems, such as the MaaS system considered in this study.}

For the simplicity of notations, we temporarily drop the index $d$ in this section. 
Mathematically, travelers with the same destination share a single MDP defined by a tuple 
$(\mathcal{S}, \A, P, u, \gamma)$, where 
$\mathcal{S} = \mathcal{V}$ and $\A = \mathcal{V}$ are the sets of nodes (\textit{states}) and links (\textit{actions}) in the multi-modal network, respectively; 
$P:\mathcal{S} \times \A \rightarrow p(\mathcal{S})$ specifies the state transition dictated by the link choice probabilities; 
$u\in\mathbb{R}^{|\mathcal{S}||\A|}$ represents the link utilities, and $\gamma \in (0, 1]$ is the discount factor.
We use $\A_i$ to denote the set of available links at node $i\in\mathcal{V}$ in the multi-modal network and $\Delta_i \coloneqq \Delta(\mathcal{A}_i)  = \{\pi(\cdot|i)|\sum_{j\in\A_i} \pi(j|i) =1; \pi(j|i) \geq 0,\forall j\in \A_i\}$ to denote the probability simplex of $\mathcal{A}_i$, where $\pi(\cdot|i)$ denotes the link choice probabilities at node $i$. 
In what follows, we will use the notations of current state $s$ and node $i$ interchangeably. For instance, 
 $\pi=(\pi(\cdot|s)_{s\in\mathcal{S}})$ will be used to represent link choice probabilities, also known as policy.  

We adopt the perturbed utility Markov choice model (PUMCM) developed in \cite{yaoperturbed} to model mode and route choices in the multi-modal network, or equivalently, the policy $\pi$ defined above. 
Following the classic MDP literature, PUMCM defines the value function $V:\mathcal{S}\rightarrow \mathbb{R}$ as the expected cumulative utility from a given state to the destination.
Differently, it introduces a state-dependent perturbation function $F_s$ to capture the unknown stochasticity in choice behaviors. By definition, $F_s$ is an essentially smooth and essentially convex function~\citep[see e.g., Ch. 12 in][]{Rockafellar+1970} of the choice probability $\pi(\cdot|s)\in \interior{(\Delta_s)}$, where $\interior(\Delta_s)$ denotes the interior of $\Delta_s$ and $\Pi$ denote the set of interior policies.

Let $V=(V(s))_{s\in\mathcal{S}}$ denote the vector of values. PUMCM defines the Bellman operator $T_\pi$ for a given policy $\pi$ as
\begin{equation}
    (T_\pi V) (s) = \mathbb{E}_{a \sim \pi(\cdot | s)}\left[u_{s,a} + \gamma \mathbb{E}_{s'\sim P(\cdot|s,a)}[V(s')]\right] - F_s(\pi(\cdot | s)),
\end{equation}
It can be proved that, under the reachability condition of policy $\pi$, there exists a unique fixed point $V_\pi \in \mathbb{R}^{|\mathcal{S}|}$ satisfying $T_\pi V_\pi = V_\pi$~\citep[Ch. 7.2 in][]{bertsekas2012dynamic, yaoperturbed}.
Further, the Bellman optimality operator $T_*$, defined as
\begin{equation}\label{eq:PUM_MDP}
(T_* V)(s) = \max_{\pi(\cdot | s) \in \interior(\Delta_s)}\; \mathbb{E}_{a \sim \pi(\cdot | s)}\left[u_{s,a} + \gamma \mathbb{E}_{s'\sim P(\cdot|s,a)}[V(s')]\right] - F_s(\pi(\cdot | s))
\end{equation}
also induces a unique fixed point under mild conditions. This result is proved in \cite{yaoperturbed} and formally stated in the following lemma. 

\begin{lemma}\label{lemma:unique_fixed_point}
    Suppose that i) the perturbation functions $F_s(\cdot), \forall s \in \mathcal{S}$ are essentially smooth and essentially convex, and ii) $T_{\pi} \mathbf{0}$ are bounded and non-positive. Then, there exists a unique fixed point $V^* \in \mathbb{R}^{|\mathcal{S}|}$ such that:
    \begin{equation}
        V^* = T_* V^*.
    \end{equation}
    In addition, the fixed point $V^*$ is optimal in the sense that $V^*\geq V_\pi, \forall \pi \in \Pi$.
\end{lemma}

While the first condition in Lemma \ref{lemma:unique_fixed_point} requires a particular design of perturbation functions, the second easily holds in real practice in multi-modal transportation systems. It essentially implies that travelers have no incentive to keep looping in the network. 
This is naturally satisfied if link utilities are negative and specified as follows:
\begin{align}
\text{MaaS:  } \quad u_{i,ij} &=\begin{cases}
    -t_{ij},\; \forall (i,j) \in \E \\
    -p_{od},\; \forall i=o \in \N_O, (o,j) \in \E^{od}_{\text{MaaS}}
\end{cases}, \label{eq:MaaS_link_utility}\\
\text{non-MaaS:  } \quad  {u}_{i, ij} &= -t_{ij} - \tilde{p}_{ij}, \; \forall (i,j) \in \tilde{\E}, \label{eq:non-MaaS_link_utility}
\\
\text{Driving:   }\quad u_{i,ij} &=\begin{cases}
    -t_{ij}- \check{p}_{ij},\; \forall (i,j) \in \check{\mathcal{E}}\\
    -\check{p}_{fix},\; \forall i=o \in \N_O, (o,j) \in \E^{od}_{\text{Drive}}
\end{cases}, \label{eq:Driv_link_utility}
\end{align}
where $p_{od}$ denotes the OD-based MaaS trip fare imposed on the dummy links starting from trip origin $o$; $t_{ij}$ refers to either travel, access or transfer times, and $\tilde{p}_{ij}$ can be either the single-modal trip fare or a transfer penalty, depending on the type of link $(i,j)$; and $\check{p}_{ij}$ denote the variable driving costs and $\check{p}_{fix}$ denote the fixed cost associated with car ownerships. {For simplicity, we set the value of time to be \$1 per minute for all travelers, while the model is general for any specification of value of time.
}



Given the link utilities $u \coloneqq \left(u_{s,a}(t, p, \tilde{p})\right)_{s \in \mathcal{S}, a \in \mathcal{A}_s}$, travelers solve the PUMCM problem as per Eq.~\eqref{eq:PUM_MDP}. It then follows immediately from Lemma~\ref{lemma:unique_fixed_point} that the corresponding link flows can be efficiently computed using $V^*$. This result is formally stated in the following lemma.

\begin{lemma}[\cite{yaoperturbed}]\label{lemma:optimal_flows}
    Under the conditions in Lemma~\ref{lemma:unique_fixed_point} and with link utilities $u$ defined in~\eqref{eq:MaaS_link_utility}-\eqref{eq:Driv_link_utility}, the optimal value function $V^*(u)$ exists and is continuously differentiable. In addition, the function of optimal link flows is derived as
    \begin{align}
        x^{*}(u) &= q \nabla V^*(u),\label{eq:optimal_demand_response}
    \end{align}
    where $q \in \mathbb{R}^{|\mathcal{S}|}$ is the demand vector from each state towards the termination state.
\end{lemma}

Let $\pi^*(u)$ denote the optimal policy corresponding to the optimal value $V^*(u)$. Lemma~\ref{lemma:optimal_flows} also leads to the following correspondence: 
\begin{equation}\label{eq:optimal_link_flow}
    x^{*}(u) = q[\mathbb{I} - \gamma \pi^*(u) P]^{-1} \pi^*(u),
\end{equation}
where invertibility is ensured by the property of PUMCM.
Further, for the entropic perturbation function $F_s$ (e.g., scaled Shannon entropy), the optimal policy is available in closed form:
\begin{align}\label{eq:optimal_policy}
    \pi^*(u) = \nabla_Q F^*(Q(u)),
\end{align}
where $Q(u) = u + \gamma P V^*(u)$ and $F^*$ is the convex conjugate~\citep[see e.g.,][]{Rockafellar+1970} of perturbation function $F$, that is twice continuous differentiable for PUMCM~\citep{yaoperturbed}. 

Moreover, \cite{yaoperturbed} derive the closed-form sensitivity of optimal link flow $x^*$ with respect to utility $u$, a key input for the leader's decision making. Specifically, for a given demand vector $q$ and a utility vector $u$, the sensitivity is available analytically as
\begin{align}\label{eq:optimal_demand_response_sensitivity}
    \nabla x^*(u) = \pi^*(u)^\top \nabla N(u) + \text{diag}\left(\Lambda N(u)\right) \nabla \pi^*(u),
\end{align}
where $N: \mathbb{R}^{|\mathcal{A}|} \rightarrow \mathbb{R}_+^{|\mathcal{S}|\times 1}$ denote the expected total number of visits to each state $s \in \mathcal{S}$; $\Lambda\in \{0,1\}^{|\mathcal{A}|\times |\mathcal{S}|}$ denote the action-state incident matrix, where $\Lambda_{a, s} = 1,\forall s\in \mathcal{S}, a \in \mathcal{A}_s$, and zero, otherwise; and 
$\text{diag}(\cdot)$ puts the entry into a diagonal matrix. Each component in Eq.~\eqref{eq:optimal_demand_response_sensitivity} also has the following closed-form expressions: 
\begin{subequations}\label{eq:optimal_demand_response_sensitivity_components}
    \begin{align}
            N(u) &= [\mathbb{I} - \gamma P^\top \pi^*(u)^\top]^{-1}  q^\top \\
            \nabla \pi^*(u)&= \nabla_Q^2 F^*(Q(u)) \left[\mathbb{I} + \gamma P [\mathbb{I} - \gamma \pi^*(u) P]^{-1} \pi^*(u) \right]\\            
            \nabla N(u) &= [\mathbb{I} - \gamma P^\top \pi^*(u)^\top]^{-1} P^\top \text{diag}(\Lambda N(u)) \nabla \pi^*(u).
    \end{align}
\end{subequations}
Note that, for utility $u$ satisfying conditions in Lemma~\ref{lemma:unique_fixed_point}, the sensitivity $\nabla x^*(u)$ is continuous in $u$.


In sum, given a set of link utilities $u$, we can uniquely identify the corresponding optimal values $V^*(u)$, from which both the routing policies $\pi^*(u)$, the resulting link flows $x^*(u)$, and their sensitivities $\nabla x^*(u)$, can be directly computed using Eqs.~\eqref{eq:optimal_demand_response}-\eqref{eq:optimal_demand_response_sensitivity_components}. 
These followers' responses are then passed to the leaders to update their decisions.

\subsection{Leader: Traffic operator} \label{sec:traffic operator}


Inspired by the dual formulation of Markovian traffic equilibrium~\citep{baillon2008markovian}, we formulate the equivalent optimization problem of the PUMCM-based traffic equilibrium, denoted as PUME hereafter. 
Let $q^d, V^{d*}$ and $x^{d*}$ denote the demands, optimal values, and link flows corresponding to destination $d\in\N_D$, respectively; and $\check{\mathcal{L}}$ denote the set of links related to the traffic equilibrium problem. Given the trip fares $(p,\tilde{p})$, the dual formulation of PUME is given by
\begin{align}\label{eq:dual-traffic-equilibrium}
    \min_{t \in \Omega_t} \; \sum_{a \in \check{\mathcal{L}}} \int_{t_{0,a}}^{t_a} z_a(v){d}v + \sum_{d\in N_D} q^d V^{d*}(t, p, \tilde{p}),
\end{align}
where $\Omega_t$ denotes the feasible set of travel times $t$, and $z_a(t_a): \mathbb{R} \rightarrow \mathbb{R}$ is a continuous function that maps from link travel time to link flow, often known as the inverse link performance function~\citep{baillon2008markovian}.  

Problem~\eqref{eq:dual-traffic-equilibrium} directly motivates the design of a \textit{virtual} traffic operator, whose primary goal is to balance traffic supply and demand in the multi-modal network. Hence, in what follows, we refer to $z_a$ as the \textit{supply function}. 
Before specifying the supply functions, we first introduce the standard boundedness assumption on travel times. 
\begin{assumption}~\label{assumption:compact_t_set}
    The feasible set of travel times is defined as $\Omega_t \coloneqq \{t| t_0 \leq t \leq M_t \mathbf{1}\}$, where $t_0 = (t_{0, a})_{a\in\check{\mathcal{L}}}$ denote the free-flow travel times, and $M_t \in \mathbb{R}$ are sufficiently large values. 
\end{assumption}

We introduce two types of links in the multi-modal network: i) congestible links that include regular road links and access links with flow-dependent access time, and ii) capacitated links that include public transit (PT) links and dummy links. 
Their supply functions are specified as follows: 
\begin{align}\label{eq:supply_function}
z_a(t_a)=\begin{cases}
    \tau_a^{-1}(t_a) \quad \quad \text{if congestible},\\
    \kappa_a \quad \quad \quad \quad \text{otherwise}
\end{cases}
\end{align}
where $\tau_a^{-1}(\cdot)$ denotes the inverse of some monotonically increasing link performance function (e.g., Bureau of Public Roads (BPR) function), and $\kappa_a$ denotes the capacity.
For all congestible links, we set $\tau_a^{-1}(t_{0,a}) = 0$  for free-flow travel time $t_{0,a}$, and the capacity of dummy links is set to infinity. For capacitated links, $t_{0,a}$ denotes the fixed travel time.

Let $t_a^\star$ denote the optimal solution to Problem~\eqref{eq:dual-traffic-equilibrium}, the optimality condition reads
\begin{equation}
    z_a(t^\star_a) + \sum_{d \in \mathcal{N}_D} q^d \nabla_{u} V^{d*}(u) \nabla_{t_a} u(t^\star, p, \tilde{p}) - \lambda^\star_a + \mu^\star_a= z_a(t^\star_a) - x_a(t^\star, p, \tilde{p}) - \lambda^\star_a + \mu^\star_a = 0,
\end{equation}
where $x^*_a(t^\star, p, \tilde{p})$ represents the aggregate MaaS and non-MaaS link flows on links $a \in \check{\mathcal{L}}$, computed as $x^*_a(t^\star, p, \tilde{p})=\sum_{d \in \mathcal{N}_D} \sum_{a' \in \mathcal{A}} x_{a'}^{d*}(t^\star, p, \tilde{p}) \delta_{a,a'}$ with $\delta_{a,a'}=1$ if the utility of action $a' \in \A$ is defined according to the travel time on link $a$, and zero, otherwise\footnote[3]{We use ``$\star$'' to denote the optimal solutions to the leader's problems so as to distinguish them from the optimal solutions to PUMCM.};
$\lambda^\star_a, \mu^\star_a$ are the dual variables associated with the lower and upper bounds of $t_a$, respectively. 

With properly selected $M_t$ (i.e., the maximum possible link travel time), 
we can ensure $t^\star_a < M_t$ and thus $\mu^\star_a = 0,\forall a\in\check{\mathcal{L}}$. The optimality condition then reduces to 
\begin{equation}~\label{eq:traffic_equilibrium_cond}
     z_a(t^\star_a) - x^*_a(t^\star, p, \tilde{p}) - \lambda^\star_a = 0.
\end{equation} 
{The following proposition specifies the values of $t^\star_a$ and $\lambda^\star_a$ at the traffic equilibrium.

\begin{proposition}\label{prop:dual}
    Suppose the link travel flows follow PUMCM and the inverse link performance function $\tau_a^{-1}(t_a)$ is monotonically increasing, then the following results hold: 
    \begin{itemize}
        \item For a congestible link $a$, it must hold that $t^\star_a> t_{0,a},\lambda^\star_a =0$.
        \item For a capacitated link $a$, if $x^*_a(t_0, p, \tilde{p}) < \kappa_a$, then $t^\star_a = t_{0,a},\lambda^\star_a > 0$; otherwise, $t^\star_a> t_{0,a},\lambda^\star_a = 0$. 
    \end{itemize}
\end{proposition}
\begin{proof}
    We first prove the result of congestible links. Due to PUMCM, the optimal link flows $x_a^{d*}(t_0, p, \tilde{p})$ is strictly positive. Besides, we have $\tau_a^{-1}(t_{0,a}) = 0$, i.e., free-flow travel time is achieved with zero flow. 
    Thus, we have $\tau_a^{-1}(t_{0,a}) - x^*_a(t_{0}, p, \tilde{p}) < 0$. Since $\tau_a^{-1}(t_a)$ is increasing with $t_a$ while $x_a^{d*}(t, p, \tilde{p})$ is non-increasing in $t_a$, we have the difference $\tau_a^{-1}(t_a) - x^*_a(t, p, \tilde{p})$ monotonically increases with $t_a$. Therefore, the optimality condition must hold at some $t_a^\star > t_{0, a}$, which yields $\lambda^\star_a = 0$. 

    For the capacitated links, we first consider the case $x^*_a(t_0, p, \tilde{p}) < \kappa_a$, i.e., with the minimum link travel time, the link travel flow is still below the capacity. It is thus easy to imply that $t^\star_a = t_{0,a}$ and $\lambda^\star_a = \kappa_a - x^*_a(t_0, p, \tilde{p}) > 0$. 
    On the other hand, if $x^*_a(t_{0}, p, \tilde{p}) \geq \kappa_a$, to ensure the capacity constraint, it must hold that $t^\star_a > t_{0, a}$ and thus $\lambda_a^\star = 0$.
\end{proof}

Proposition~\ref{prop:dual} implies that the congestible link supply $z_a(t^\star_a)$ must link flow $x^*_a(t^\star, p, \tilde{p})$ at the equilibrium travel time $t_{0, a} < t_a^\star < M_t$. Meanwhile, when the capacitated link is fully occupied, the difference $t^\star_a-t_{0,a}$ reflects the congestion effect.}

Since the feasible set $\Omega_t$ is compact convex and the objective~\eqref{eq:dual-traffic-equilibrium} is convex, it must hold that the equilibrium travel times $t^\star$ are equivalent to the solutions of the following VI conditions~\citep{nagurney2013network}: 

For a given $(p,\tilde{p})$, find $t^\star\in\Omega_t$ such that
\begin{equation}\label{eq:traffic_VI}
\Big\langle z(t^\star) - x^*(t^\star, p, \tilde{p}),\; t - t^\star \Big\rangle  
\geq 0, \quad \forall t\in \Omega_t,
\end{equation}
where, $x^*(t, p, \tilde{p}) \coloneqq (x_a^*(t, p, \tilde{p}))_{a \in \mathcal{E} \setminus \mathcal{E}_{\text{dummy}}}$. Note that the traffic equilibrium problem involves not only physical links (e.g., road and PT links) but also access links that reflect the waiting times for different services (e.g., waiting and boarding times for bus, waiting and pickup times for on-demand mobility service). 

\subsection{Leader: MaaS platform} \label{sec:MaaS}


Recall that the MaaS platform makes payments to service operators based on the consumed service capacity.
Let $c_m \in \mathbb{R}^{|\E_m|}$ denote the link-specific unit wholesale capacity price the operator $m$, and the set of wholesale prices as $c \coloneqq \{ c_m \}_{m \in \mathcal{M}}$. Then, for a given $(t, c, \tilde{p})$, the MaaS platform's optimal pricing problem is formulated as 
\begin{equation}\label{eq:MaaS_opt_v0}
    \max_{p \in \Omega_p} \; p^\top Q(t, p, \tilde{p}) - \sum_{m\in\M} c_m^\top X_m(t, p, \tilde{p}),
\end{equation}
where $\Omega_p$ is the feasible set for MaaS fares, $Q(t, p, \tilde{p}) = \left(\sum_{a \in \E^{od}_\text{MaaS}}x_a^{d*}(u(t, p, \tilde{p}))\right)_{o \in \mathcal{N}_o, d \in \mathcal{N}_d}$ is the MaaS demand, and $X_m(t, p, \tilde{p}) = \left(\sum_{d\in\N_D}x_a^{d*}(u(t, p, \tilde{p}))\right)_{a\in \E_m}$ is the travel flows on the MaaS subnetwork $G_m$.

To safely transform Problem~\eqref{eq:MaaS_opt_v0} into a VI problem, we make the following assumption.
\begin{assumption}\label{assumption:bounded_p}
There exists a price upper bound $M_p \in \mathbb{R}$ such that, at the upper bound $M_p$, elasticities of MaaS demands $Q(\cdot, M_p\mathbf{1}, \cdot)$ and $X_m(\cdot, \cdot, M_p\mathbf{1}), \forall m \in \mathcal{M}$ are (approximately) zero .
\end{assumption}
\noindent This assumption implies that, at the price upper bound $M_p$, MaaS demand $Q(\cdot, M_p\mathbf{1}, \cdot)$ remains at approximately zero, and $X_m(\cdot, \cdot, M_p\mathbf{1}), \forall m \in \mathcal{M}$ remains at some constant, regardless of a further increase in price $p$ and $\tilde{p}$, respectively. 
Correspondingly, we define a compact convex feasible set $\Omega_p = \{p | 0 \leq p \leq M_p\mathbf{1} \}$ and formulate the corresponding VI problem: 

For a given $(t, c, \tilde{p})$, find $p^\star \in \Omega_p$ satisfying
\begin{align}\label{eq:MaaS_VI}
    \left\langle c^\top \nabla_p X(t, p^\star, \tilde{p}) - p^{\star\top} \nabla_{p} Q(t, p^\star, \tilde{p}) - Q(t, p^\star, \tilde{p}), \; p - p^\star \right\rangle \geq 0, \quad \forall p \in \Omega_p.
\end{align}
Note that Problems~\eqref{eq:MaaS_opt_v0} and \eqref{eq:MaaS_VI} are not equivalent, due to the non-convexity of Problems~\eqref{eq:MaaS_opt_v0}. Nevertheless, all solutions to Problems~\eqref{eq:MaaS_opt_v0} satisfy the VI condition \eqref{eq:MaaS_VI}. Besides, the VI conditions also imply that, at the optimal solution $p^\star$, the marginal cost $c^\top \nabla_p X(t, p^\star, \tilde{p})$ should equal the marginal revenue $p^{\star\top} \nabla_{p} Q(t, p^\star, \tilde{p}) + Q(t, p^\star, \tilde{p})$.

\subsection{Leaders: Mobility service operators} \label{sec: mobility operator}

Since we assume the service capacities are fixed and exogenous, the optimal pricing problem for each service operator reduces to maximizing its total revenue, which consists of the trip fare revenue from its own service and the payment from the MaaS platform. 
Let $\tilde{p}_{-m}$ denote the trip fares of mobility services other than $m\in\M$. For a given $(t, p, \tilde{p}_{-m})$, the optimal pricing problem for each operator $m\in\M$ is formulated as 
\begin{subequations}\label{eq:TNC_opt}
    \begin{align}
    \max_{\tilde{p}_{m} \in \Omega_{\tilde{p}_m}} \; & \tilde{p}^\top_m \tilde{X}_m(t, p, \tilde{p}_m, \tilde{p}_{-m}) + c_m^\top X_m(t, p, \tilde{p}_m, \tilde{p}_{-m}) \label{eq:TNC_MaaS_coupling}
\end{align}
\end{subequations}
where $\Omega_{\tilde{p}_m}$ is a compact and convex feasible set defined as $\Omega_{\tilde{p}_m} = \{\tilde{p}_{m} | 0 \leq \tilde{p}_{m} \leq M_p\mathbf{1} \}$, and  $\tilde{X}_m(t, p, \tilde{p}_m, \tilde{p}_{-m}) \coloneqq \left(\sum_{d \in \N_D}x_a^{d*}(u(t, p, \tilde{p})\right)_{a \in \tilde{\E}_m}$ is the travel flows on the non-MaaS subnetwork $\tilde{G}_m$.

The VI problem for each service operator is thus defined as:

For a given $(t, p, \tilde{p}_{-m})$, find $\tilde{p}^\star_{m} \in  \Omega_{\tilde{p}_m}$ such that
\begin{align}\label{eq:operator_VI}
\langle - \tilde{p}^{\star\top}_m \nabla_{\tilde{p}_m} \tilde{X}_m(t, p, \tilde{p}^\star_m, \tilde{p}_{-m}) &- \tilde{X}_m(t, p, \tilde{p}^\star_m, \tilde{p}_{-m}) \nonumber \\
& - c_m^\top \nabla_{\tilde{p}_m} X_m(t, p, \tilde{p}^\star_m, \tilde{p}_{-m}),\; \tilde{p}_m - \tilde{p}^\star_m \rangle  
 \geq 0, \; \forall \tilde{p}_m \in \Omega_{\tilde{p}_m}.
\end{align}
Similar to the MaaS platform, the VI condition~\eqref{eq:operator_VI} states that, at the optimal solution, the marginal revenue generated from non-MaaS trips cancels out the marginal revenue generated from MaaS trips. 
This is due to the modeling assumption and the property of PUMCM.
As travelers do not move across MaaS, non-MaaS, and driving subnetworks, these three modes are strict substitutes. Consequently, an increase in non-MaaS price  
necessitates a decrease in non-MaaS demand and an increase in MaaS and driving demands. In other words, an increase in non-MaaS price brings about a loss in non-MaaS revenue and a gain in MaaS revenue. 

\section{Equilibrium and solution approach}\label{sec:equilibrium}
\subsection{Equilibrium as a single-level VI problem}

With all leaders' and followers' problems defined in previous sections, we are ready to establish the equilibrium condition for the multi-leader-multi-follower game as a single-level VI problem.  


For a given wholesale price $c\coloneqq(c_m)_{m \in \mathcal{M}}$, the variational Nash equilibrium (VNE)~\citep{facchinei2003finite} of the multi-leader-multi-follower game is a state $(t^\star, p^\star, \tilde{p}^\star) \in \Omega$ that simultaneously satisfies
\begin{align}\label{eq:GNE_VI_orig}
    & \Big\langle z(t^\star) - x^*(t^\star, p^\star, \tilde{p}^\star), t - t^\star \Big\rangle \nonumber\\
    & \quad + \sum_{m\in\mathcal{M}} \left\langle - \tilde{p}^{\star\top}_m \nabla_{\tilde{p}_m} \tilde{X}_m(t^\star, p^\star, \tilde{p}^\star_m, \tilde{p}^\star_{-m}) - \tilde{X}_m(t^\star, p^\star, \tilde{p}^\star_m, \tilde{p}^\star_{-m}) - c^\top_m \nabla_{\tilde{p}_m} X_m(t^\star, p^\star, \tilde{p}^\star_m, \tilde{p}^\star_{-m}), \tilde{p}_m - \tilde{p}^\star_m \right\rangle \nonumber \\
    &\quad + \left\langle c^{\top} \nabla_p X(t^\star, p^\star, \tilde{p}^\star) - p^{\star\top} \nabla_{p} Q(t^\star, p^\star, \tilde{p}^\star) - Q(t^\star, p^\star, \tilde{p}^\star), p - p^\star \right\rangle    \geq 0, 
    \quad  \forall (t, p, \tilde{p})\in \Omega, 
\end{align}
where $\Omega \coloneqq \Omega_t \times \Omega_{{p}} \times \Omega_{\tilde{p}}$ is the joint feasible set, $x^{d*}$ is the optimal link flows of PUMCM given $(t, p, \tilde{p})\in \Omega$, and $x^*, X, Q, \tilde{X}$ are induced by $x^{d*}$ as specified in previous sections. 
Define $y = \text{col}(t, p, \tilde{p})$, then VNE \eqref{eq:GNE_VI_orig} is represented as VI$(G, \Omega)$:
\begin{align}\label{eq:GNE_VI}
    \text{Find $y^\star\in \Omega$ such that $\langle G(y^\star), y-y^\star\rangle\geq0,\;\forall y\in \Omega.$}
\end{align}


\begin{proposition}[VNE existence] \label{prop:existence}
Suppose the Assumptions~\ref{assumption:compact_t_set}-~\ref{assumption:bounded_p} hold.
Then, the multi-leader-multi-follower game admits a VNE as the solution to VI$(G, \Omega)$.
\end{proposition}
\begin{proof}
Since optimal value function $V^*(u): \mathbb{R}_{-}^{|\mathcal{A}|} \rightarrow \mathbb{R}^{\mathcal{S}}$ is twice continuous differentiable with non-positive utility $u$~\citep{yaoperturbed} and $u:y\mapsto \mathbb{R}_{-}^{|\mathcal{A}|}$ is negative, we have that $G(y)$ is continuous on the set $\Omega$. Further, $\Omega$ is the Cartesian product of multiple convex compact sets, which is also convex and compact. Then, by Theorem 1 in~\cite{nagurney2008variational}, VI$(G, \Omega)$ admits a solution. 
\end{proof}

Proposition~\ref{prop:existence} implies that, for any wholesale price $c$, there exists an equilibrium for the multi-leader-multi-follower game. 
Consequently, the strategic decision on wholesale price can be viewed as an equilibrium selection or screening process. 
{This is because the wholesale capacity price primarily governs how revenue is allocated between the MaaS platform and service operators. Different admissible price-setting rules would lead to different wholesale prices, while each of them induces a particular VNE.}
Given the scope of this paper, we do not further analytically investigate this problem but numerically explore the impact of $c$ via a sensitivity analysis in Section~\ref{sec:experiment-small-capacity}.
{It is also worth noting that the equilibrium defined in~\eqref{eq:GNE_VI_orig} is not unique in general, as the map $G$ is typically non-monotone on $\Omega$. However, it is easy to ensure local uniqueness with the additional assumption that $G$ is locally strictly monotone in an open neighborhood of an equilibrium solution. This would then allow a consistent interpretation of counterfactual analysis based on a given baseline equilibrium.
We will numerically demonstrate the existence of local uniqueness in Section~\ref{sec:convergence}.
}

\subsection{Solution algorithm}\label{sec:solution}
Since VI$(G, \Omega)$ is non-monotone, the classical methods that require strictly monotone mapping, such as the project method and general iterative scheme~\citep{dafermos1983iterative}, cannot guarantee a satisfactory convergence performance. Hence, we instead adopt the modified project method, also known as extragradient \citep{korpelevich1976extragradient}, which demonstrates better performance in our numerical experiments. 

The algorithm is rooted at the equivalent transformation from VI to a fixed point~\citep{nagurney2008variational}. For any parameter $\gamma >0$, VI$(G, \Omega)$ is equivalent to the following fixed point problem:
\begin{align}~\label{eq:proj_alg}
    y^\star = \text{Proj}_{\Omega}(y^\star - \gamma G(y^\star)),
\end{align}

Instead of simple projection, the Korpelevich's extragradient method performs a ``prediction-correction'' procedure at every iteration $k$:
\begin{align}
    y^{(k+1/2)} &= \text{Proj}_{\Omega}(y^{(k)} - \gamma G(y^{(k)})) \\
    y^{(k+1)} &= \text{Proj}_{\Omega}(y^{(k)} - \gamma G(y^{(k+1/2)})).
\end{align}
Although the extragradient method is only proven to be convergent for (pseudo)monotone VIs, it often shows satisfactory convergence performance for non-monotone VIs with a suitable step size. In our numerical experiments, we also test different step sizes and choose the one that yields the best convergence. 

The extragradient method tackles the update of pricing schemes and travel times, which requires gradients evaluated on the optimal link flows of travelers $x^*$. This requires solving the PUMCM problem to obtain the optimal values $V^*$. 
Fortunately, it has been shown in \cite{yaoperturbed} that {simple value iteration can already achieve global convergence with asymptotic linear rate when the conditions in Lemma~\ref{lemma:unique_fixed_point} are satisfied}. 
Additionally, the sensitivities of $x^*$ with respect to the link utilities, i.e., $\nabla_u x^*(u)$, have analytical form and thus can be computed explicitly. 
The readers are referred to \cite{yaoperturbed} for details about the solution algorithm of PUMCM and the derivation of sensitivities.

Algorithm~\ref{algo} summarizes the solution procedure of the proposed MaaS model. In brief, at each iteration, we first update link utilities $u$ based on the current travel times and pricing strategies $(t, p, \tilde{p})$. Then the lower-level PUMCM is solved in parallel to obtain the link flows $x^*(u)$ and sensitivities $\nabla x^*(u)$ for all OD pairs simultaneously, with which the mapping $G$ is constructed for the extragradient update that also allows for parallel updates. Note that the lower-level PUMCM problem is solved twice in each iteration.

\begin{algorithm}[H] \label{algo}
\DontPrintSemicolon
\KwIn{Multi-modal network $\mathcal{G}$, OD demand $q$, wholesale price $c_m$}

\textbf{Parameters:} Step sizes $\gamma$, gap threshold $\epsilon$, residual function $R(\cdot)$, maximum number of iterations $N$.\;
\textbf{Initialize:} $y^{(0)}$.\;

\For{$k = 0, 1, 2, \dots, N-1$}{
    \textbf{1. Prediction step}
    
    1.1. Compute link utilities $u^{(k)} = u(y^{(k)})$ using Eqs.~\eqref{eq:MaaS_link_utility}-\eqref{eq:Driv_link_utility}.\;

    1.2. Solve lower-level PUMCM and obtain $x^{*(k)}$ using Eq.~\eqref{eq:optimal_link_flow}.\;

    1.3. Evaluate $G(y^{(k)}))$ using flows $x^{*(k)}$ and flow sensitivities $\nabla x^{*(k)}$ using Eq.~\eqref{eq:optimal_demand_response_sensitivity}. 
    
    1.4. Prediction \;
    \Indp
    $y^{(k+1/2)} = P_\Omega(y^{(k)} - \gamma G(y^{(k)}))$\;
    \Indm

    \textbf{2. Correction step}

    2.1-2.3 Repeat steps 1.1-1.3 for $y^{(k+1/2)}$

    2.4 Correction \;
    \Indp
    $y^{(k+1)} = P_\Omega(y^{(k)} - \gamma G(y^{(k+1/2)}))$\;
    \Indm

    \textbf{3. Check convergence}\;
    \Indp
    \If{
    $R(y^{(k+1)}) < \epsilon$\;
}{
    break\;
    \Indm
}
}

\caption{Algorithm for VI Problem}
\end{algorithm}


{As a direct consequence of the VNE formulation in~\eqref{eq:GNE_VI_orig}, the projection steps 1.4 and 2.4 in Algorithm~\ref{algo} are computationally inexpensive, since they only involve box constraints. 
The main computation then reduces to evaluating the operator $G$, which can be efficiently handled within the PUMCM framework by exploiting the sparsity of the underlying network structure; see~\cite{yaoperturbed} for implementation details. 
This stands in contrast to classical bilevel formulations, where each operator evaluation would require solving a full traffic equilibrium problem, a difficulty avoided here through the introduction of a ``virtual'' traffic operator.}

\section{Numerical experiments}\label{sec:experiment}

To validate the proposed model and solution algorithm, as well as to generate insights into MaaS systems with strategic operators, we design two sets of numerical experiments. The first one is based on a small network and focuses on analyzing the trade-offs of the MaaS platform and service operators subject to various transfer penalties and wholesale capacity prices. 
The second set of experiments is designed based on the Sioux Falls network and aims to demonstrate the scalability of the proposed solution algorithm meanwhile providing insights into the real-world implementation of MaaS. 

The multi-modal networks used in both experiments are constructed according to Section~\ref{sec:network}, while the setup of different types of links is detailed below. 
All algorithms are implemented in Python and PyTorch 2.1. The small network experiments are executed on a MacBook with M2 chip, and the Sioux Falls experiments are performed on a HPC with V100 GPU. We make use of MPS/GPU acceleration for PyTorch when available.

\subsection{Small network}\label{sec:illustrative}

\subsubsection{Setup}\label{sec:experiment-small-setup}
\begin{figure}[htb]
  \centering
  \includegraphics[width=0.8\textwidth]{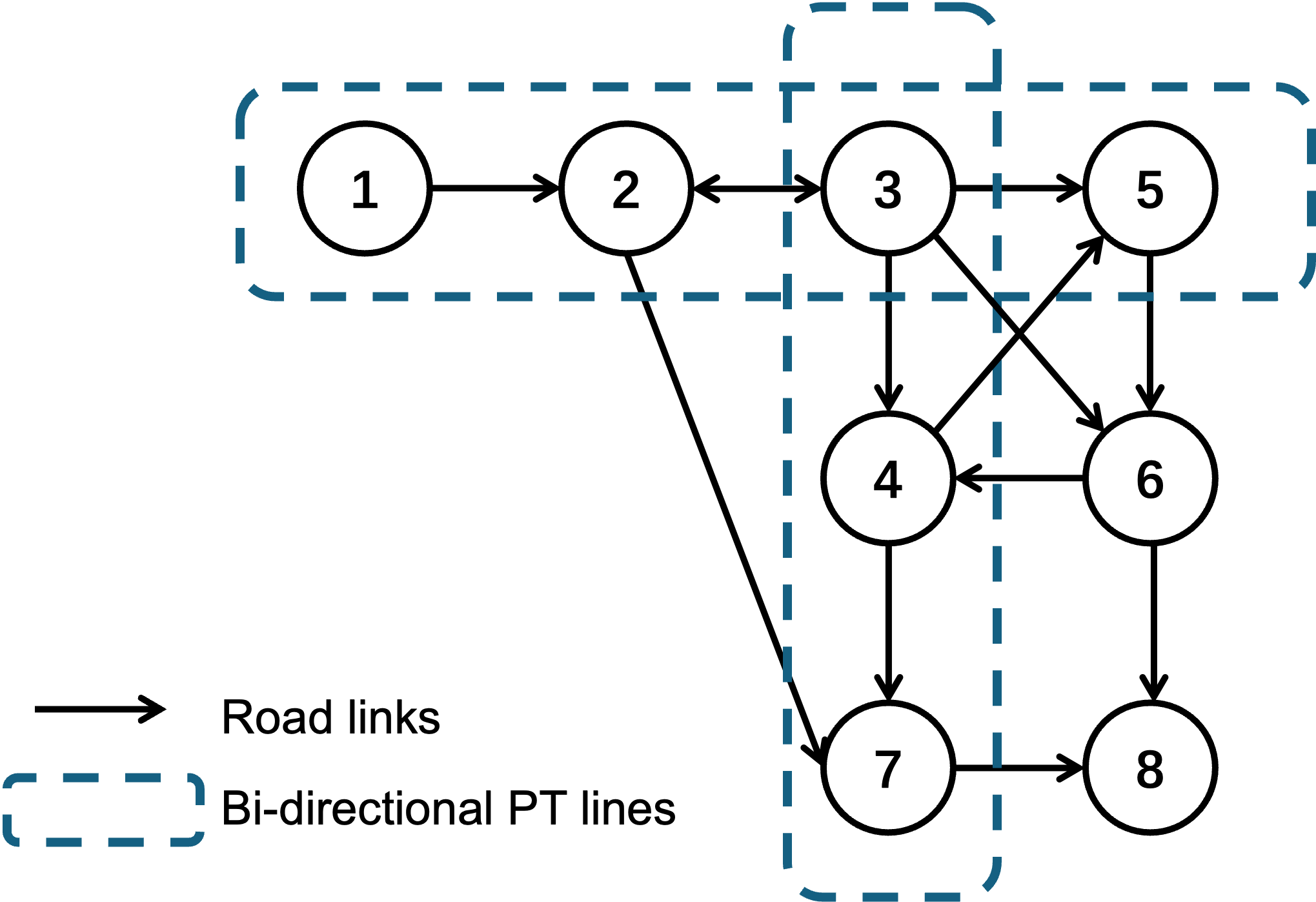} 
  \caption{Small network.}
  \label{fig:small network}
\end{figure}

The network used in this experiment consists of 8 physical nodes and 11 {road links, which represent the physical road infrastructure shared by private driving and mobility-on-demand (MoD) services}. In addition, two PT lines are designed that span over 3 and 4 physical nodes, respectively, and intersect at Node 3. {The road links and PT links are modeled as separate and independent components of the network: driving and MoD demands jointly induce congestion on road links, whereas PT links operate with fixed travel times and are subject to capacity constraints.}
For simplicity, we assume both PT lines are operated by a single PT operator and follow the same pricing scheme. 
We also assume there is a single MoD service in the system that operates on all {road links} and supports inter-modal transfers at all PT stops, while transfers only occur at the same physical node. Similarly, the intra-modal transfer between the two PT lines only occurs at Node 3. 

We apply the Bureau of Public Roads (BPR) function to model the travel time on the {road links}:
\begin{align}\label{eq:bpr}
    t_a(x_a) = t_{0, a}\left[1 + \alpha \left(\frac{x_a}{\kappa_a}\right)^\beta\right],
\end{align}
where $x_a$ is the link flow that aggregates the MoD (both MaaS and non-MaaS) and private driving, $\kappa_a$ is link capacity (or saturation flow), and $\alpha, \beta$ are parameters. 

The BPR function is also used to model the access time of the MoD service in the corresponding access links. Specifically, we assume the access time is homogeneous across the network and is determined by the total MoD travel demand and vehicle fleet size. The former is computed by aggregating all access link flows to the MoD subnetwork, while the latter is fixed as per the assumption. They are then treated as link flow $x_a$ and capacity $\kappa_a$ in Eq.~\eqref{eq:bpr}, respectively. 
On the other hand, PT links have fixed travel time but limited capacity, as the capacitated link defined in Section~\ref{sec:traffic operator}. The waiting time is counted in the access links to the PT subnetwork and assumed to be constant as well.

Table~\ref{tab:O-D pair} reports the demand profile designed in this experiment. We consider three OD pairs with total demands of 1000. The first and second OD pairs share the same destination, while the second and third OD pairs share the same origin. Travelers between the first two OD pairs can take a direct PT trip to their destinations, whereas those traveling Node 1 to Node 7 must make a transfer at Node 3 if they choose to take transit.

\begin{table}[H]
    \centering
    \begin{tabular}{c|c|c}
    \toprule
       \textbf{Origin} & \textbf{Destination} & \textbf{Demand}\\
       \midrule
        3 & 5 & 500\\
        1 & 5 & 300\\
        1 & 7 & 200\\
        \bottomrule
    \end{tabular}
    \caption{Demand profile of small network.}
    \label{tab:O-D pair}
\end{table}

Recall that the MaaS price is OD-based (see Eq.~\eqref{eq:MaaS_link_utility}). As for the non-MaaS services, we assume the MoD price is distance-based, whereas the PT price is computed by the number of PT link traveled. Accordingly, the non-MaaS link price is specified as 
\begin{align}
    &\tilde{p}_{ij}= \begin{cases} \label{non-MaaS prices}
        \tilde{p}^M l_{ij}, \quad \forall(i,j)\in \tilde{\E}_{MoD} \\
        \tilde{p}^P, \quad \forall(i,j)\in \tilde{\E}_{PT},
    \end{cases}
\end{align}
where $l_{ij}$ refers to the link length, and $\tilde{p}^M,\tilde{p}^P$ are the unit prices for MoD and PT services, respectively, which are also the operators' pricing decisions. 


As for private driving, we assume the total cost is computed as a fixed cost (e.g., parking) plus a distance-based variable cost (e.g., fuel expenses). The former is assigned to the access link to the driving subnetwork, while the latter is added to each {road link} in the driving subnetwork. 
\begin{align}
    &\check{p}_{ij}=\begin{cases} \label{driving prices}  \check{p} * l_{ij}, \quad \forall(i,j)\in \check{\E}\\
        \check{p}^{fix} , \quad \forall o \in \N_O, (o,j) \in {\E}^{od}_{\text{Drive}}.
    \end{cases}
\end{align}


The link parameters and other exogenous variables are summarized in Appendix.\ref{sec:appendix- setting}.

\subsubsection{{Convergence of solution algorithm and robustness of equilibrium}} \label{sec:convergence}

\begin{figure}[H]
  \centering
  \includegraphics[width=0.6\textwidth]{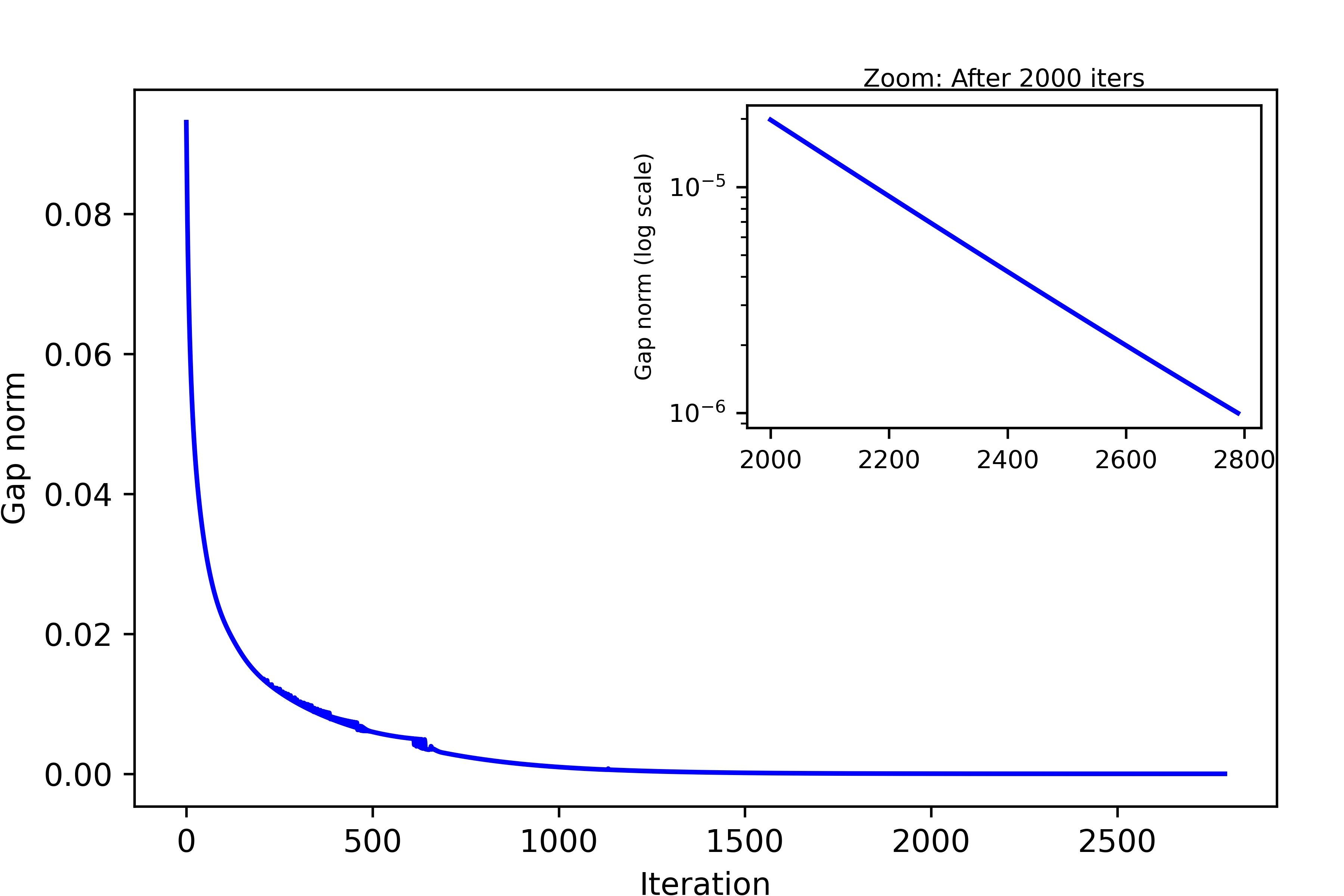} 
  \caption{Solution algorithm convergence.}
  \label{fig:convergence}
\end{figure}

In all experiments, we set the residual function to be $l_2$-norm of leaders' decision variables $(t, p, \tilde{p})$ between two consecutive iterations. 
Figure~\ref{fig:convergence} illustrates how the residual evolves over iterations. 
As can be seen, the residual decreases smoothly and quickly drops below $10^{-2}$ within 500 iterations. With a threshold $10^{-6}$, the algorithm converges after 2800 iterations. Even in the last few hundred iterations, the algorithm manages to maintain a linear rate, which demonstrates the efficacy and robustness of the extragradient method. 

{
As noted in Section~\ref{sec:solution}, the VNE~\eqref{eq:GNE_VI} is in general non-unique due to the non-monotone map $G$. However, we are able to numerically demonstrate the local uniqueness in our setting by showing that different initial solutions lead to the same equilibrium. 
Specifically, we shift the initial point $y^{(0)}$ with multiplicative perturbations of $\theta = 1\%, 5\%, 10\%$, i.e., $ y^{(0)}_{\theta,i} = (1 + \theta) y^{(0)}_i$, on some randomly selected variable $i$. The subset of perturbed variables corresponds to 20\%, 40\%, 60\%, 80\%, and 100\% of all decision variables. 
10 replications are conducted for each combination of shifting magnitude and ratio, which results in a total of 150 sample points.

\begin{figure}[htb]
  \centering
  \includegraphics[width=0.8\textwidth]{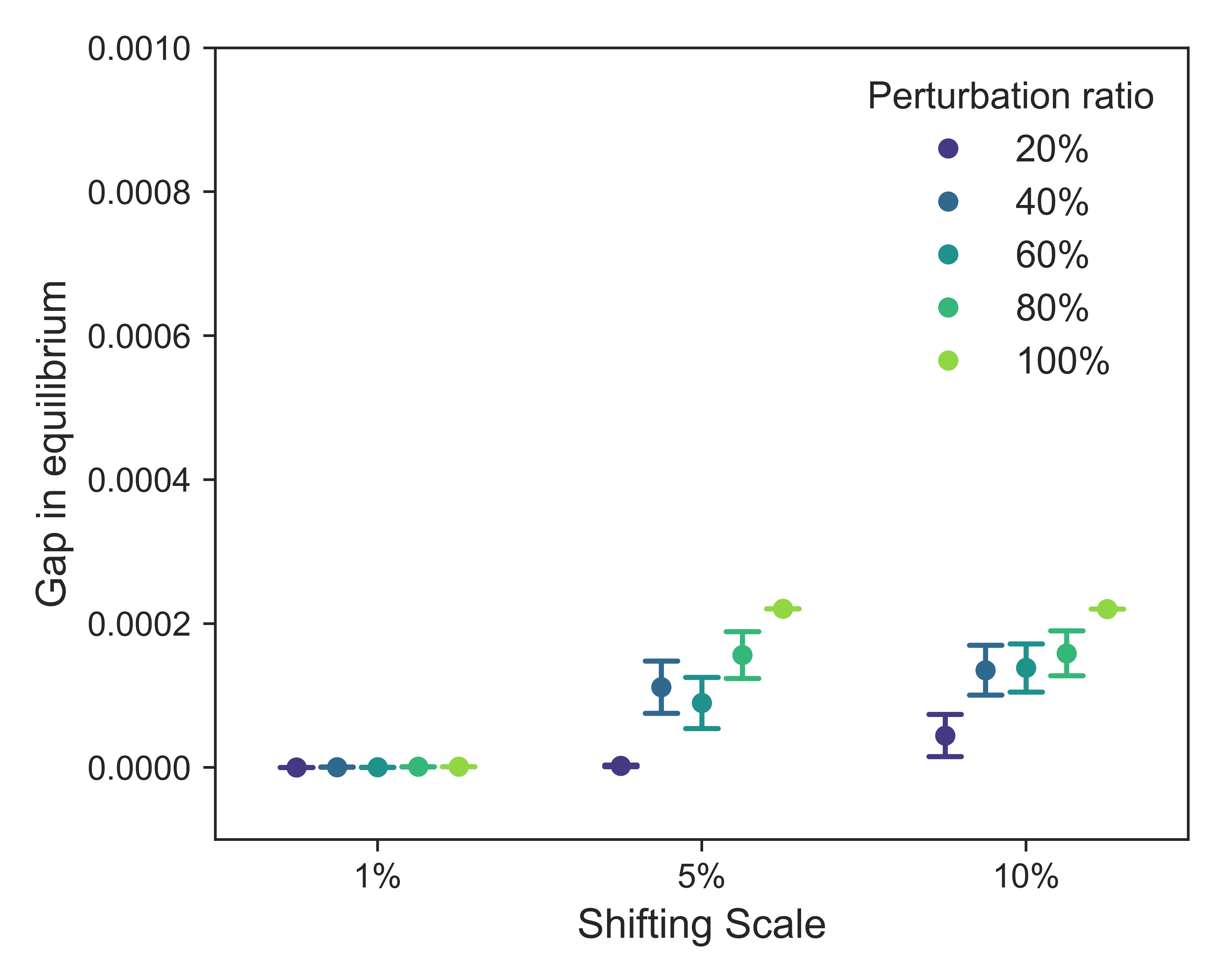} 
  \caption{Gap in equilibria due to shift of initial solutions.}
  \label{fig:Local uniqueness}
\end{figure}

Figure~\ref{fig:Local uniqueness} illustrates the gap in equilibria obtained from shifted initial points relative to the baseline equilibrium. As expected, larger shifting scales and ratios lead to greater deviations. Nevertheless, the gap, measured in terms of $\ell_2$-norm, remains very small (below $2\times10^{-4}$) and can be considered negligible. 
Moreover, the gap does not further increase when the shifting scale is raised from 
5\% to 10\%, which indicates a strong robustness of the equilibrium solution.
}

\subsubsection{System outcomes with and without MaaS}\label{sec:experiment-small-transfer}

\begin{figure}[htb]
  \centering
  \begin{subfigure}[t]{0.49\textwidth}
    \centering
    \includegraphics[width=\linewidth]{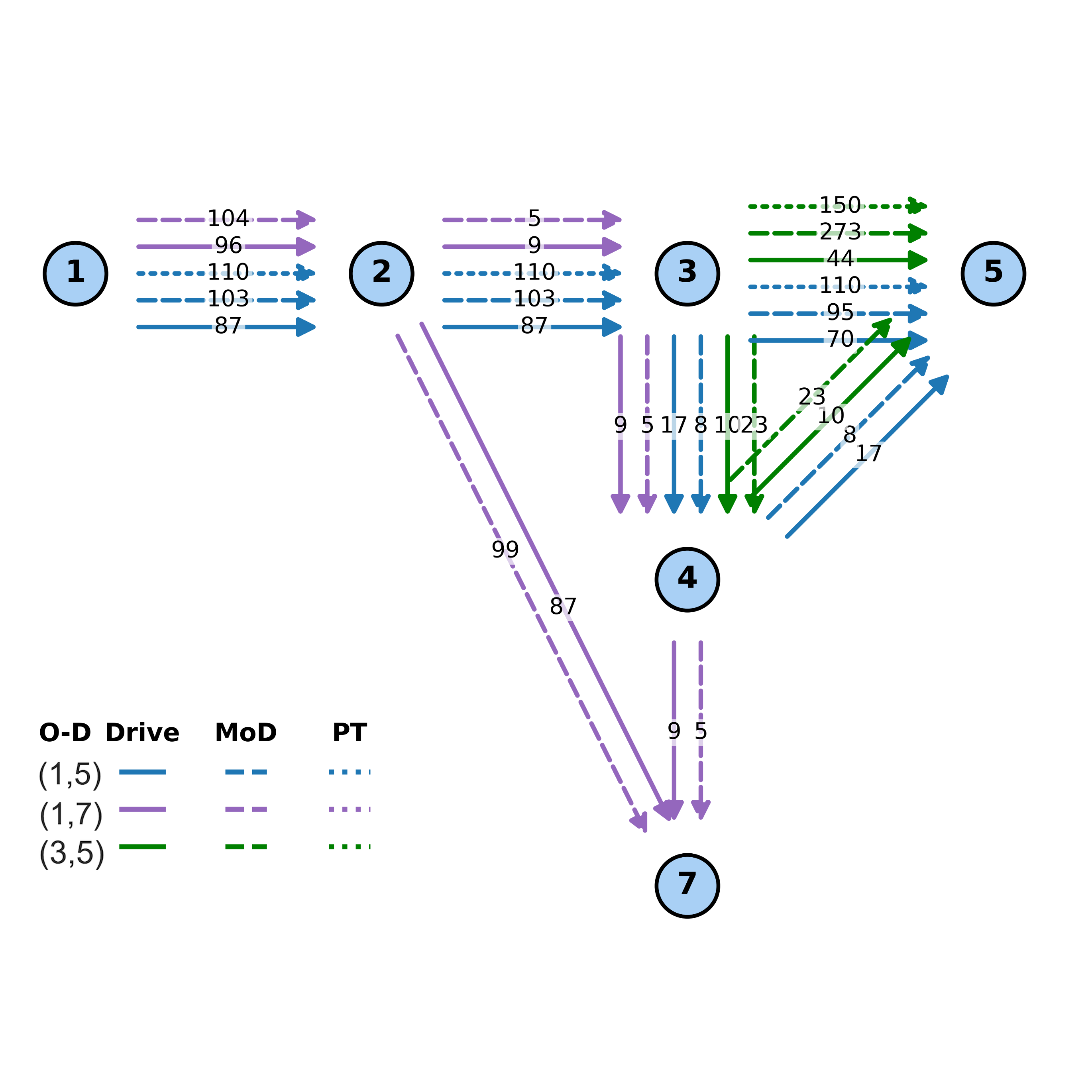}
    \caption{Without MaaS}
    \label{fig:OD flow without MaaS}
  \end{subfigure}
  \hfill 
  \begin{subfigure}[t]{0.49\textwidth}
    \centering
    \includegraphics[width=\linewidth]{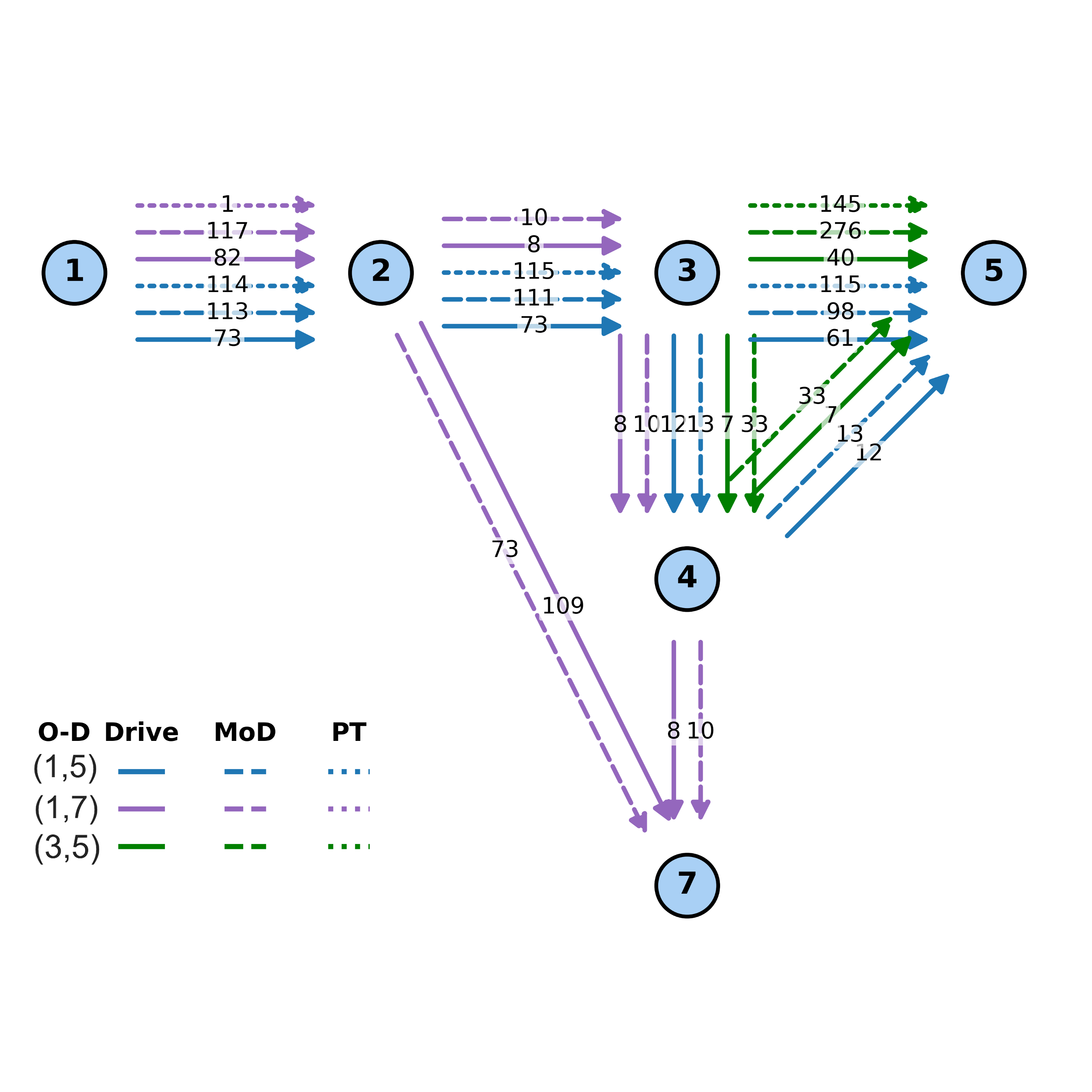}
    \caption{With MaaS}
    \label{fig:OD flow with MaaS}
  \end{subfigure}
    \caption{Equilibrium link flows of small network with and without MaaS.}\label{fig:OD_flow}
\end{figure}

{Figure~\ref{fig:OD_flow} presents the equilibrium driving, MoD, and PT link flows with and without MaaS, where flows associated with different OD pairs are distinguished by color. To simplify the presentation, links with extremely small flows are dropped. 

Overall, the introduction of MaaS does not alter the active links in the network, though it substantially redistributes demand flows across modes and links. 
In particular, a significant decrease in driving flows is observed when comparing the major {road links} (e.g., (1, 2), (3, 4), (4, 5)) after introducing MaaS. This implies that MaaS can substitute driving, especially in the most congested regions of the network.
An exception is link (2,7), where the driving flow increases from 87 to 109. This increase, however, is offset by a simultaneous decrease in the corresponding MoD flow from 99 to 73. As a result, the total vehicular flow on this link remains nearly unchanged, suggesting that MaaS primarily reallocates demand between private driving and MoD services rather than inducing additional traffic.

Figure~\ref{fig:OD_flow} also reveals clear modal shifts induced by MaaS. For OD pair (1,5), PT usage increases noticeably after the introduction of MaaS, accompanied by a shift toward MoD services. Although, as expected, MoD remains generally more attractive than PT for travelers, MaaS facilitates greater PT adoption by encouraging multi-modal trips and reducing generalized costs, thereby shifting travel demand away from private driving.}

{

We proceed to investigate the impact of transfer penalty on the system outcomes with and without MaaS.}
It is widely acknowledged that, by integrating multiple mobility services into a single on-demand service, MaaS could significantly reduce the cost of multi-modal trip planning and the inconvenience of inter- and intra-modal transfers~\citep{MARKARD2012955,jittrapirom2017mobility}. 
To differentiate transfer costs in MaaS and non-MaaS, we include a constant transfer penalty to the access links that refer to transfers. Accordingly, we solve the equilibrium model for three scenarios: i) the status quo without MaaS; ii) MaaS with a large reduction in transfer penalty, and iii) MaaS with a moderate reduction in transfer penalty. Note that case ii) is also the baseline scenario presented in Figure~\ref{fig:OD_flow}.




\begin{table}[H]
  \centering
  \small
  \setlength{\tabcolsep}{2pt} 
  \caption{Aggregate performance of small network with and without MaaS.}
  \label{tab:Comparison of results with and without MaaS}
  \begin{tabular}{lccccc}
    \toprule
    & \textbf{Without MaaS} & \multicolumn{3}{c}{\textbf{With MaaS}} \\
    Transfer penalty & 100\% & 10\% && 30\% &\\
    \midrule
    \textbf{Market share (\%)} & & & & &\\
    \quad non-MaaS & 76.3 & 58.6 && 59.4 &\\
    \quad \quad -MoD & 50.8 & 39.8 &&40.3 &\\
    \quad \quad -PT & 25.5 & 18.8  && 19.1&\\
    \quad MaaS & -- & 18.9 && 18.1 & \\
    \quad \quad -MoD & -- & 12.2 &&  11.7 &\\
    \quad \quad -PT & -- & 6.7  &&6.4 &\\
    \quad Driving & 23.7 & 22.5 &$\textcolor{red}{\downarrow\, 1.2\%}$  & 22.5\% &$\textcolor{red}{\downarrow\, 1.2\%}$\\
    
    \textbf{Transfer flow} & 0 & 15 && 1.9 &\\
    \midrule
    \textbf{Profit (\$)} & & & & &  \\
    \quad PT & 664.1 & 678.6 &$\textcolor{blue}{\uparrow\, 2.2\%}$ &  678.8 &$\textcolor{blue}{\uparrow\, 2.2\%}$  \\
    \quad MoD & 1176.7 & 1278.1 &$\textcolor{blue}{\uparrow\, 8.6\%}$ & 1276.2 &$\textcolor{blue}{\uparrow\, 8.5\%}$ \\
    \quad MaaS & -- & 232.9  &&220.6 &\\
    \quad Total & 1840.8 & 2189.6 &$\textcolor{blue}{\uparrow\, 18.9\%}$ & 2175.5 &$\textcolor{blue}{\uparrow\, 18.2\%}$\\
    \midrule
    \textbf{non-MaaS price} & & & &&\\
    \quad PT & 1.38 & 1.44  && 1.44 &\\
    \quad MoD & 0.63 & 0.66 && 0.66 &\\
    \midrule
    \textbf{Traveler welfare (\$)} & -19514 & -19287 &$\textcolor{blue}{\uparrow\, 1.16\%}$ & -19291 &$\textcolor{blue}{\uparrow\, 1.14\%}$\\
    \textbf{Social welfare (\$)} & -17673 & -17097 &$\textcolor{blue}{\uparrow\, 3.26\%}$ & -17116 &$\textcolor{blue}{\uparrow\, 3.16\%}$\\
    \bottomrule
  \end{tabular}
\end{table}

Table~\ref{tab:Comparison of results with and without MaaS} reports the main aggregate statistics of the three tested scenarios. The first observation is that the introduction of MaaS attracts more travelers to MoD and PT services, leading to a smaller market share of private driving ($-1.2\%$) in both MaaS scenarios. Meanwhile, both PT and MoD operators enjoy a higher profit, and the increases are similar in the two MaaS scenarios, i.e., around 2\% for PT and 8.5\% for MoD. Accordingly, the total profit of the MaaS platform and service operators also shares a similar growth, along with a minor improvement in traveler welfare {(i.e., the optimal value $V^*(o)$ at origin $o$, representing the expected traveler utilities at equilibrium)} and social welfare {(i.e., the sum of total traveler welfare and total operator profit). }
Overall, the introduction of MaaS is demonstrated to bring a ``win-win-win'' situation among the MaaS platform, service operators, and travelers. Yet, in the tested scenario, the service operators benefit the most, implying that the majority of surplus generated by MaaS is absorbed into the operators' profit. 

Although the overall system performance remains similar between the two MaaS scenarios, the different transfer penalties indeed result in a substantial gap in the transfer flow. When the transfer cost is largely mitigated, more transfer behaviors are observed in the system. Yet, a moderate reduction in transfer cost is sufficient to encourage multi-modal trips, which are not observed in the case without MaaS at all. 
Besides, the results indicate that the capability of reducing transfer penalties could be critical for the success of MaaS, as in our experiment, the increase in the MaaS platform's profit is largely driven by the reduction in transfer penalties.

\subsubsection{Sensitivity analysis on wholesale capacity price}\label{sec:experiment-small-capacity}

In this section, we proceed to analyze the impact of wholesale capacity price on the MaaS system. As per Proposition~\ref{prop:existence}, the multi-leader-multi-follower game in the MaaS system always admits an equilibrium. Yet, the equilibrium could vary with the wholesale capacity price. Hence, it is worth exploring if there exists a range of wholesale price $c_m$ within which the ``win-win-win'' situation observed in the previous section is guaranteed. 


For simplicity, we apply the same wholesale price to MoD and PT operators and vary it from 0 to 3.5. The zero wholesale price means the MaaS platform can consume the service capacities for free. Although this is unlikely to happen in real practice, it provides a good reference point and indicates the best-case scenario for the MaaS platform. The upper bound of the wholesale price is determined numerically. When $c_m$ increases beyond this threshold, the MaaS platform would exit the market because it cannot make a profit.

\begin{figure}[htb]
  \centering

  \begin{subfigure}[t]{0.5\textwidth}
    \centering
    \includegraphics[width=\linewidth]{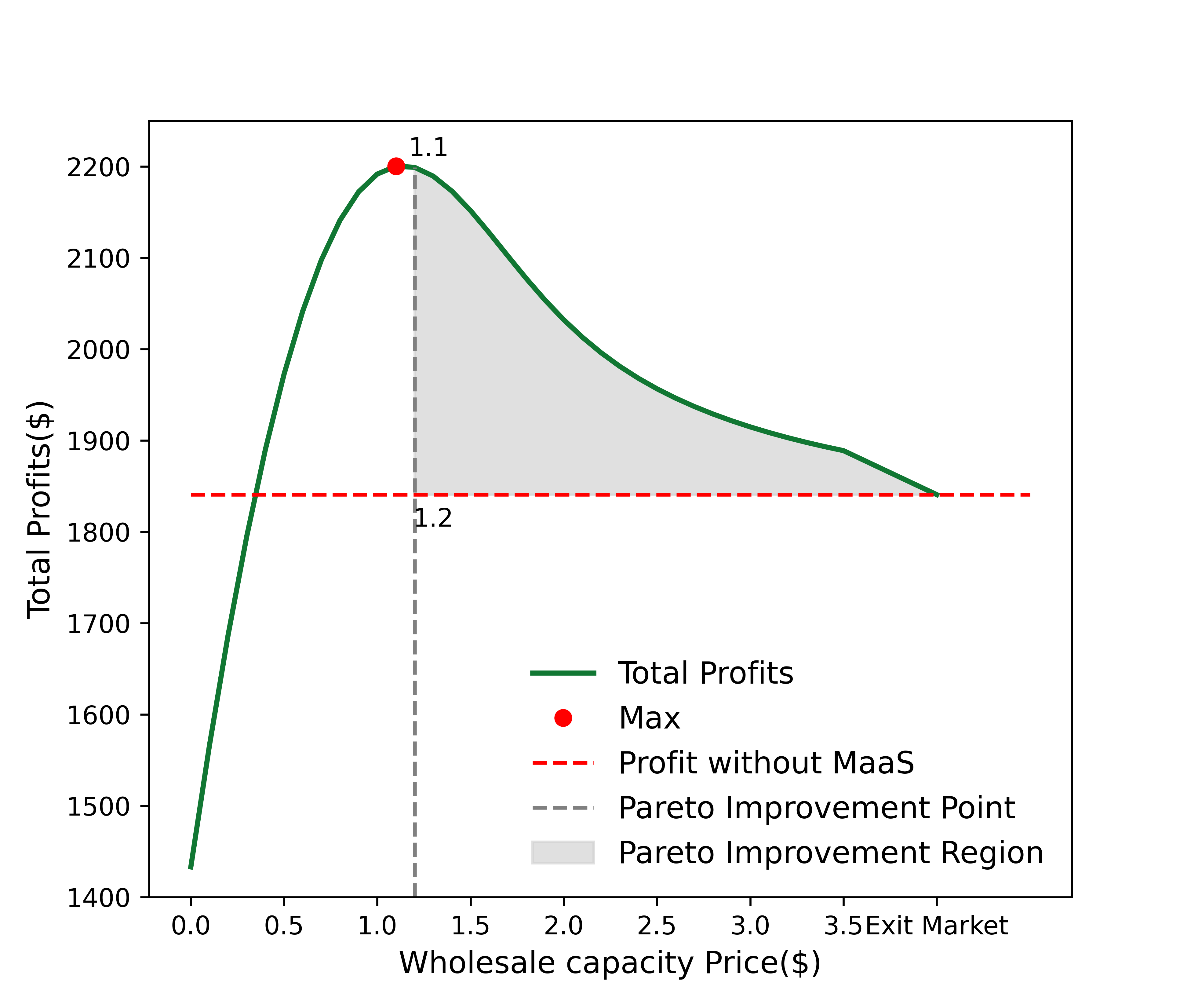}
    \caption{Total profit}
    \label{fig:total profit change}
  \end{subfigure}
  \hfill 
  \begin{subfigure}[t]{0.46\textwidth}
    \centering
    \includegraphics[width=\linewidth]{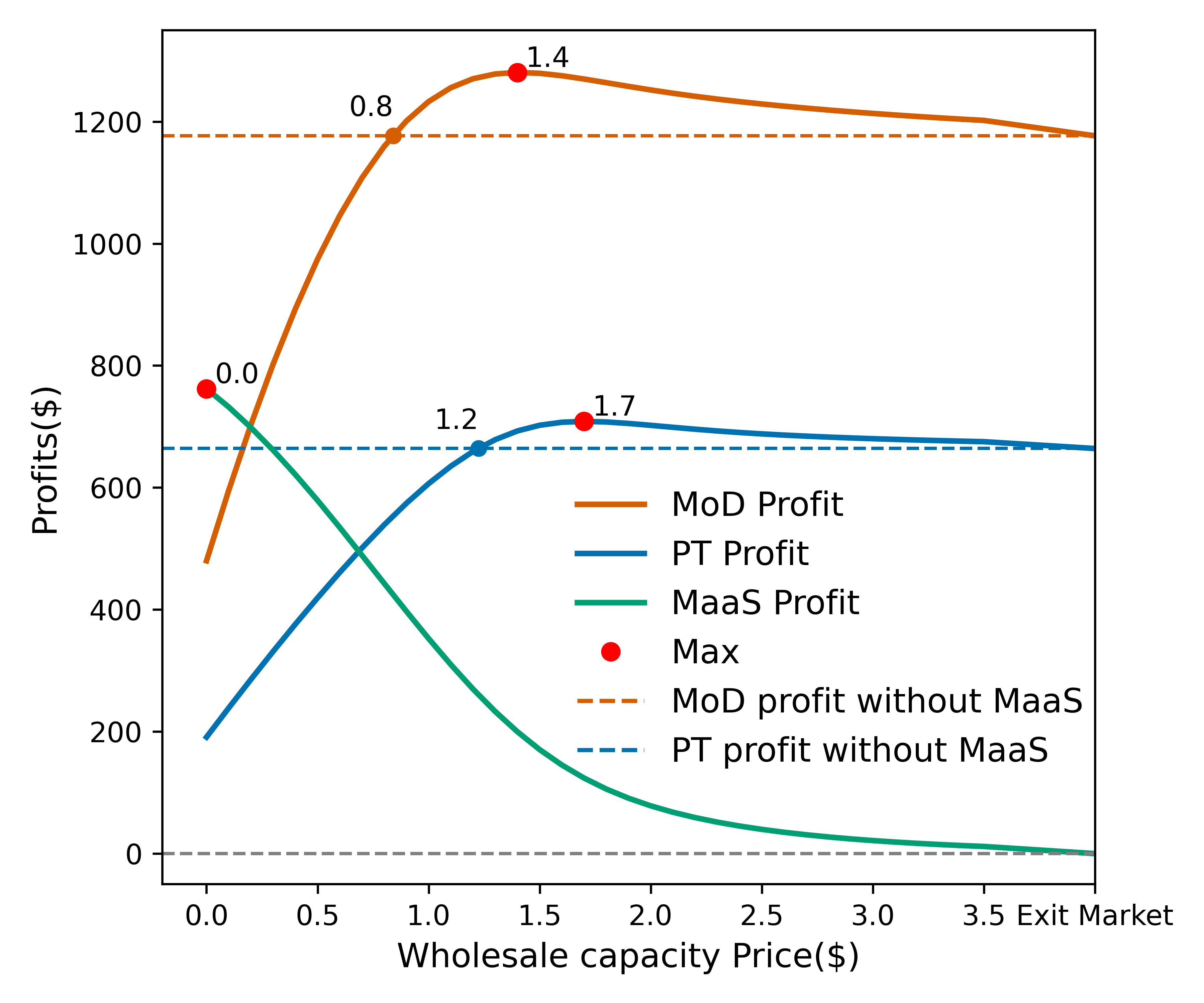}
    \caption{Service operators \& the MaaS platform}
    \label{fig:operators profits change}
  \end{subfigure}
    \caption{Total profit and profit by operator against wholesale capacity price.}
    \label{fig:profit change}
\end{figure}

Figure~\ref{fig:profit change} illustrates the variations of total profit and profit by operator with the wholesale capacity price. As shown in Figure~\ref{fig:total profit change}, as $c_m$ increases, the total profit initially rises quickly from a fairly low level, peaks at $c_m=1.1$, then decreases to the scenario without MaaS.  
A similar trend is observed for MoD and PT profits in Figure~\ref{fig:operators profits change}. However, the maxima do not coincide and the service operators achieve their largest profits at a higher wholesale price, i.e., $c_m=1.4$ for MoD and $c_m=1.7$ for PT. On the other hand, the MaaS platform reaches its maximum profit at the lower bound $c_m=0$, which is expected as in this case it does not induce any operation cost. 
As $c_m$ increases, the MaaS platform profit monotonically decreases and approaches zero. 

A more interesting observation is made when we identify the Pareto-improving regime, within which all operators enjoy a positive profit gain. This can be easily done graphically by drawing horizontal lines of PT and MoD profits in the scenario without MaaS. As shown in Figure~\ref{fig:operators profits change}, the MoD operator generates a larger profit when $c_m \geq 0.8$, while the threshold for the PT operator is slightly higher at $c_m \geq 1.2$. It thus concludes that the Pareto-improving condition is $c_m \geq 1.2$, which is marked as the grey area in Figure~\ref{fig:total profit change}. 
This result also implies how much $c_m$ should be expected in real practice. 
Since the wholesale price is negotiated between the MaaS platform and service operators, without intervention, it can only fall in the Pareto-improving regime as per the typical feasibility condition of bargaining mechanisms~\citep{thomson1994cooperative}. 
However, one can easily recognize that the wholesale price yielding maximum total profit ($c_m = 1.1$) does not lead to a Pareto improvement. In other words, additional revenue-sharing mechanisms may be necessary to achieve maximum profitability in the MaaS system. 
We leave this as a future research question. 

\begin{figure}[htb]
  \noindent
  \makebox[\textwidth][c]{
    \includegraphics[width=0.5\textwidth]{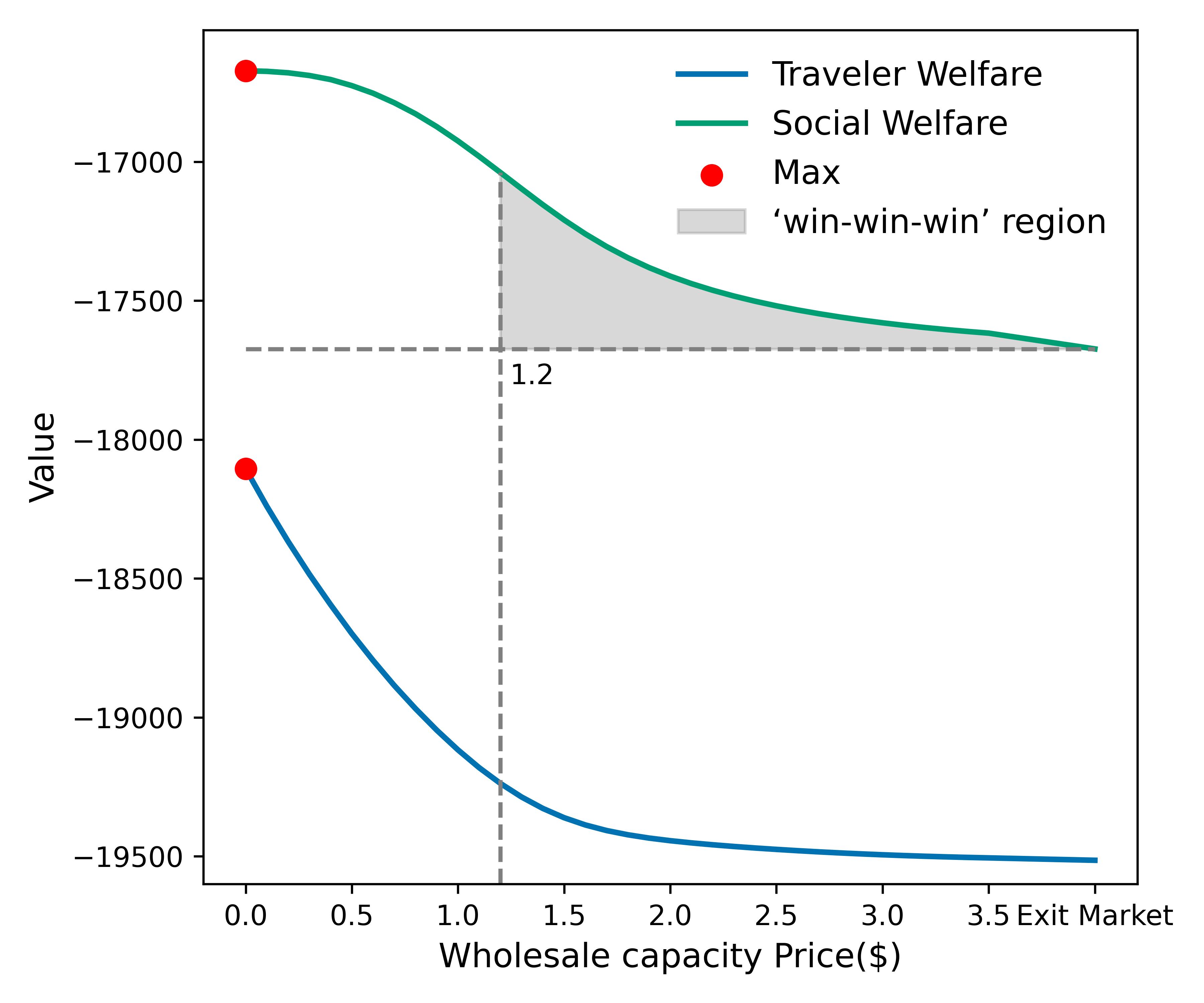}}
  \caption{System and traveler welfare against wholesale capacity price.}
  \label{fig:SW}
\end{figure}

\begin{figure}[htb]
  \centering
  \begin{subfigure}[t]{0.5\textwidth}
    \centering
    \includegraphics[width=\linewidth]{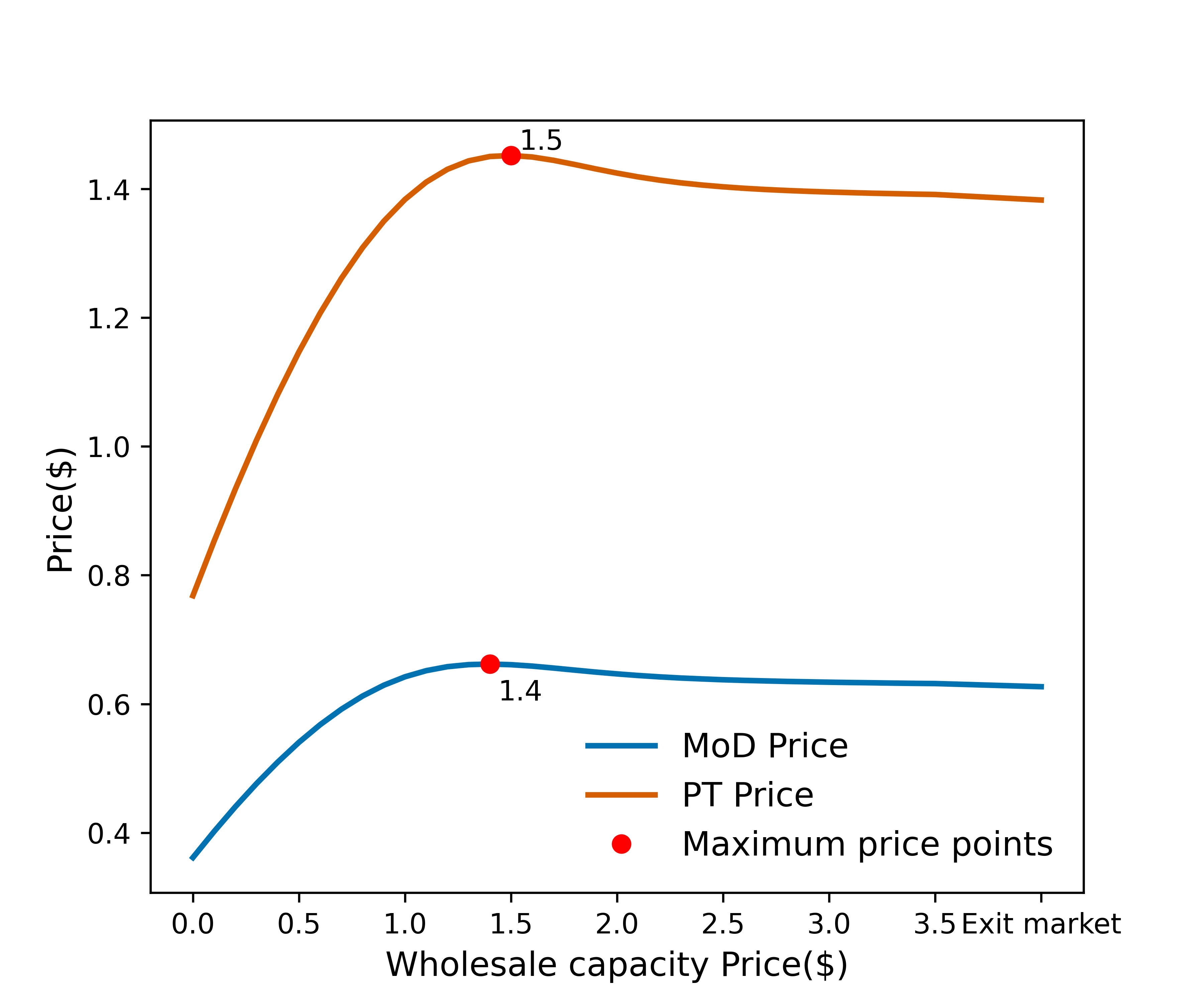}
    \caption{Non-MaaS unit price}
    \label{fig:nonmaas prices change}
  \end{subfigure}
  \hfill
  \begin{subfigure}[t]{0.46\textwidth}
    \centering
    \includegraphics[width=\linewidth]{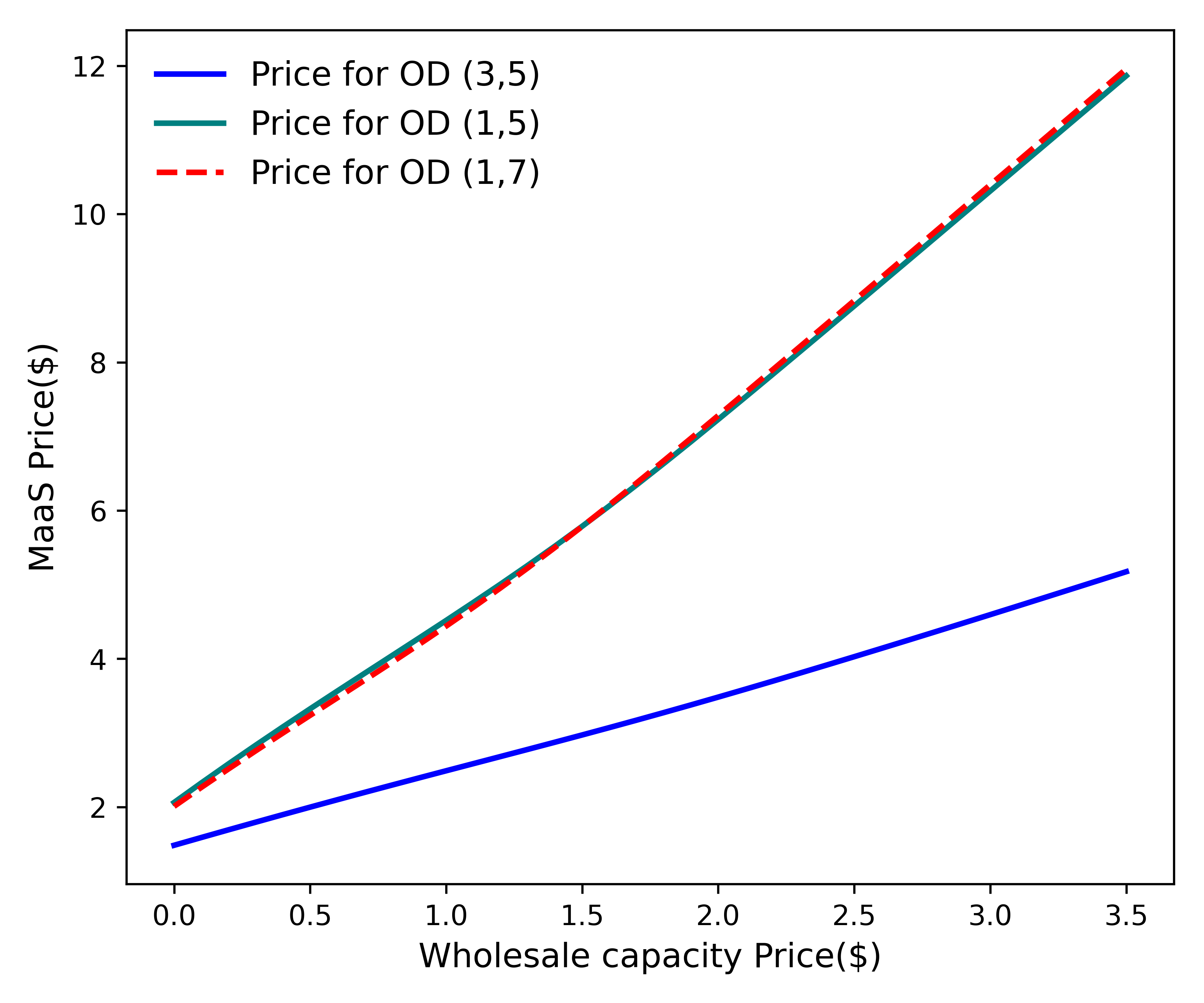}
    \caption{MaaS OD-based price}
    \label{fig:maas price change}
  \end{subfigure}

  \caption{Pricing strategy against wholesale capacity price.}
  \label{fig:price change}
\end{figure}

To complete the analysis of the condition of ``win-win-win'' situations, we plot the social and traveler welfare against the wholesale price in Figure~\ref{fig:SW} and highlight the Pareto-improving regime derived above. 
It can be found that, under the current setting, 
the introduction of MaaS is always beneficial to travelers and the entire system, and their welfare is maximized at $c_m = 0$. 
In this case, MaaS and non-MaaS travelers are both compensated with lower trip fares, thanks to the severe competition between the MaaS platform and service operators (see Figure~\ref{fig:price change}). 
However, such an ideal state is not economically stable because the improvement in traveler welfare largely comes at the cost of operator profit (see Figure~\ref{fig:total profit change}). 
Instead, the system is more likely to evolve into the gray area, where the ``win-win-win'' condition is sustained.  
Under the current setting, travelers receive a marginal benefit from the MaaS system in the Pareto-improving regime, similar to the results reported in Table~\ref{tab:Comparison of results with and without MaaS}. Yet, as will be shown in Section~\ref{sec:experiment-large}, the potential traveler improvement could largely depend on the network topology and demand profile. Hence, we leave the in-depth analysis of the existence and influential factors of Pareto-improving wholesale capacity prices to future research. 

{\subsubsection{Sensitivity analysis on travelers' value of time}

In this section, we relax the assumption of homogeneous travelers and allow travelers associated with different OD pairs to differ in their value of time. The utility functions Eqs.~\eqref{eq:MaaS_link_utility}–\eqref{eq:Driv_link_utility} are modified as follows:
\begin{align}
\text{MaaS:  } \quad u_{i,ij} &=\begin{cases}
    -v_{od}t_{ij},\; \forall (i,j) \in \E \\
    -p_{od},\; \forall i=o \in \N_O, (o,j) \in \E^{od}_{\text{MaaS}}
\end{cases}, \label{eq:newMaaS_link_utility}\\
\text{non-MaaS:  } \quad  {u}_{i, ij} &= -v_{od}t_{ij} - \tilde{p}_{ij}, \; \forall (i,j) \in \tilde{\E}, \label{eq:newnon-MaaS_link_utility}
\\
\text{Driving:   }\quad u_{i,ij} &=\begin{cases}
    -v_{od}t_{ij}- \check{p}_{ij},\; \forall (i,j) \in \check{\mathcal{E}}\\
    -\check{p}_{fix},\; \forall i=o \in \N_O, (o,j) \in \E^{od}_{\text{Drive}}
\end{cases}, \label{eq:newDriv_link_utility}
\end{align}
where $v_{od}$ denote the value of time for travelers between $(o, d)$.

\begin{figure}[htbp]
  \centering
  \begin{subfigure}[t]{0.48\textwidth}
    \centering
    \includegraphics[width=\linewidth]{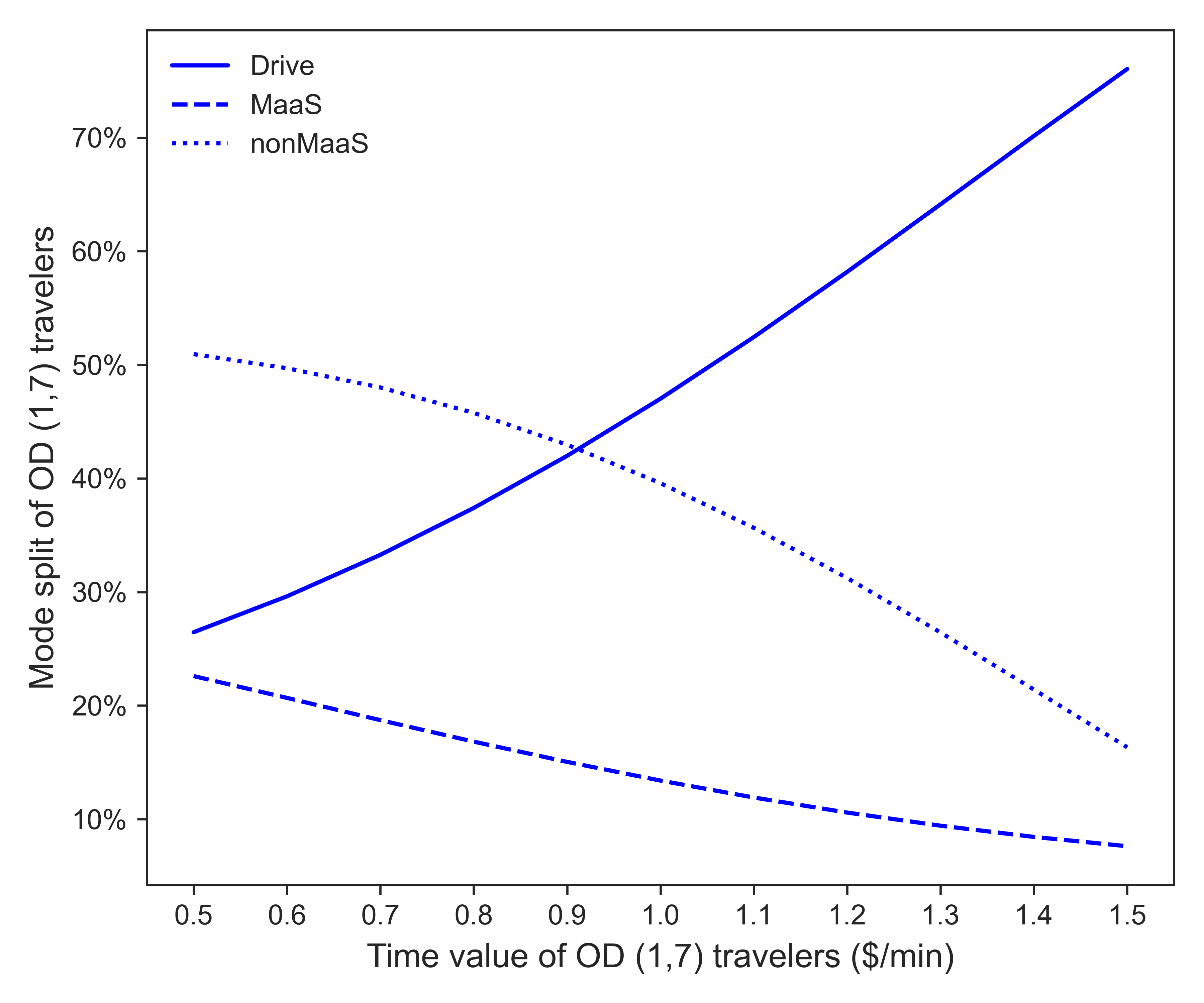}
    \caption{Modal split between OD (1,7)}
    \label{fig:VTT_sensitivity_split}
  \end{subfigure}
  \hfill
  \begin{subfigure}[t]{0.48\textwidth}
    \centering
    \includegraphics[width=\linewidth]{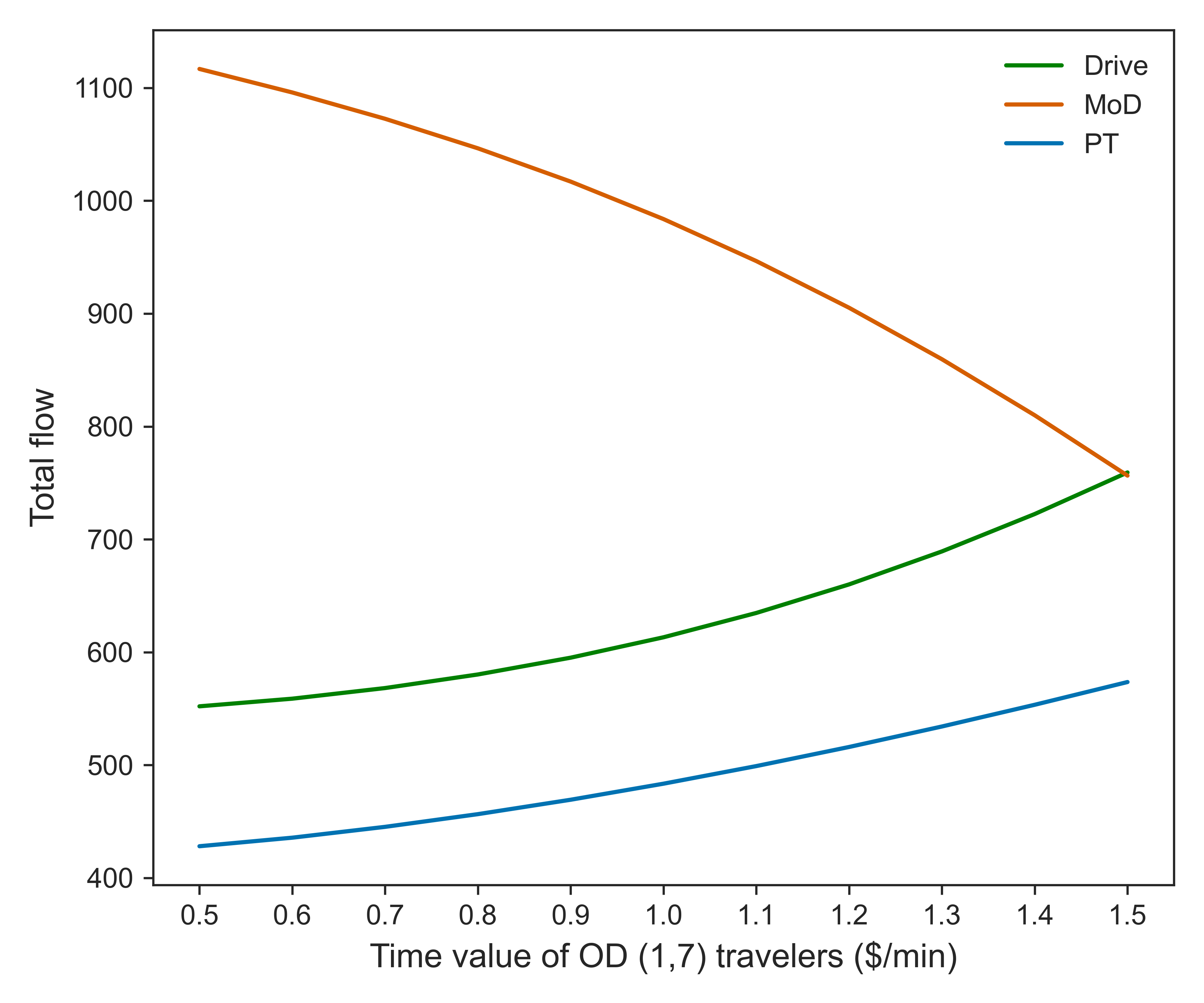}
    \caption{Mode-specific total link flows}
    \label{fig:VTT_sensitivity_flow}
  \end{subfigure}
  \hfill
  \begin{subfigure}[t]{0.48\textwidth}
    \centering
    \includegraphics[width=\linewidth]{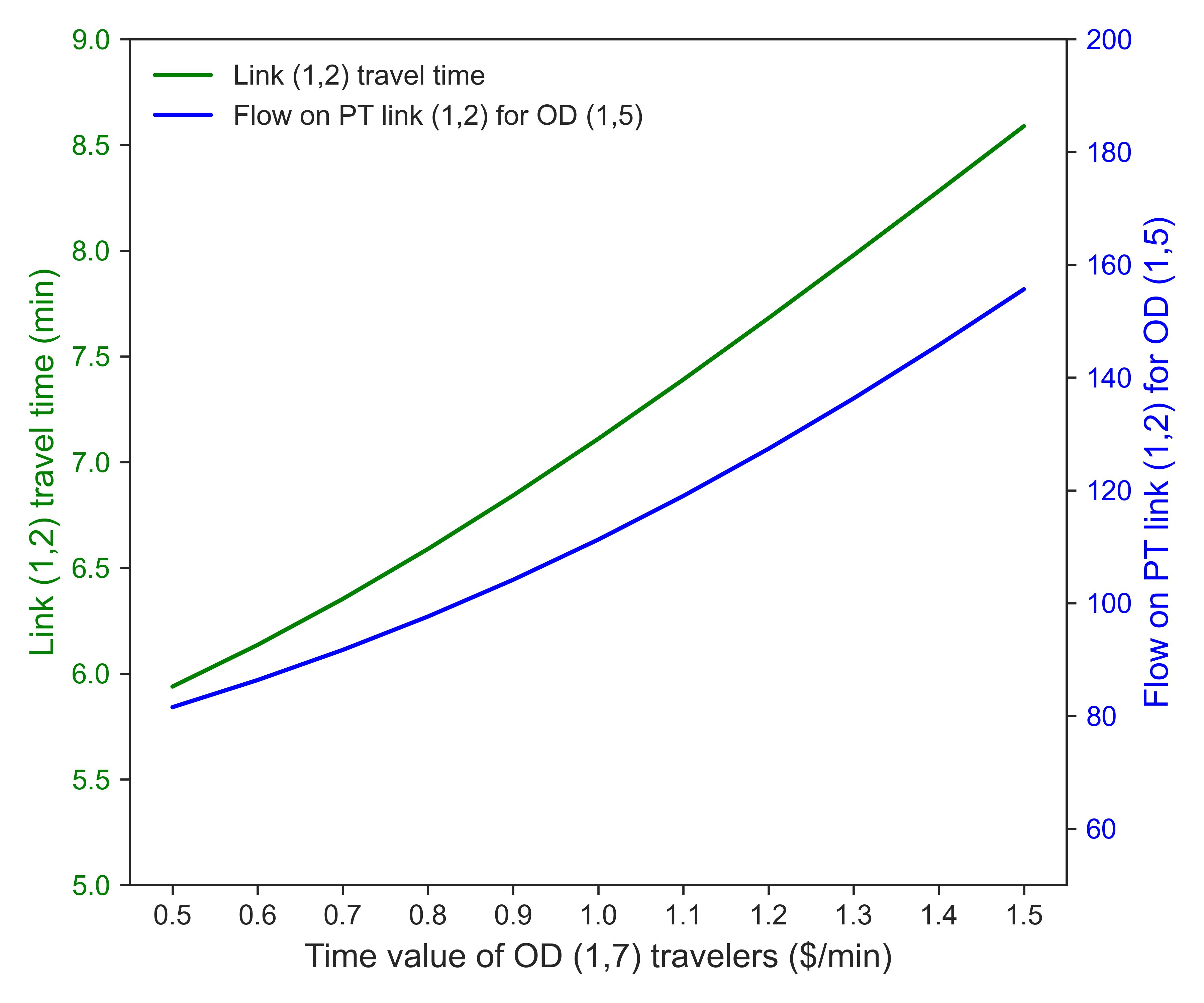}
    \caption{Travel time of road link (1,2) and travel flow on PT link (1,2) between OD pair (1,5).}
    \label{fig:VTT_sensitivity_TTPT}
  \end{subfigure}
  \hfill
  \begin{subfigure}[t]{0.48\textwidth}
    \centering
    \includegraphics[width=\linewidth]{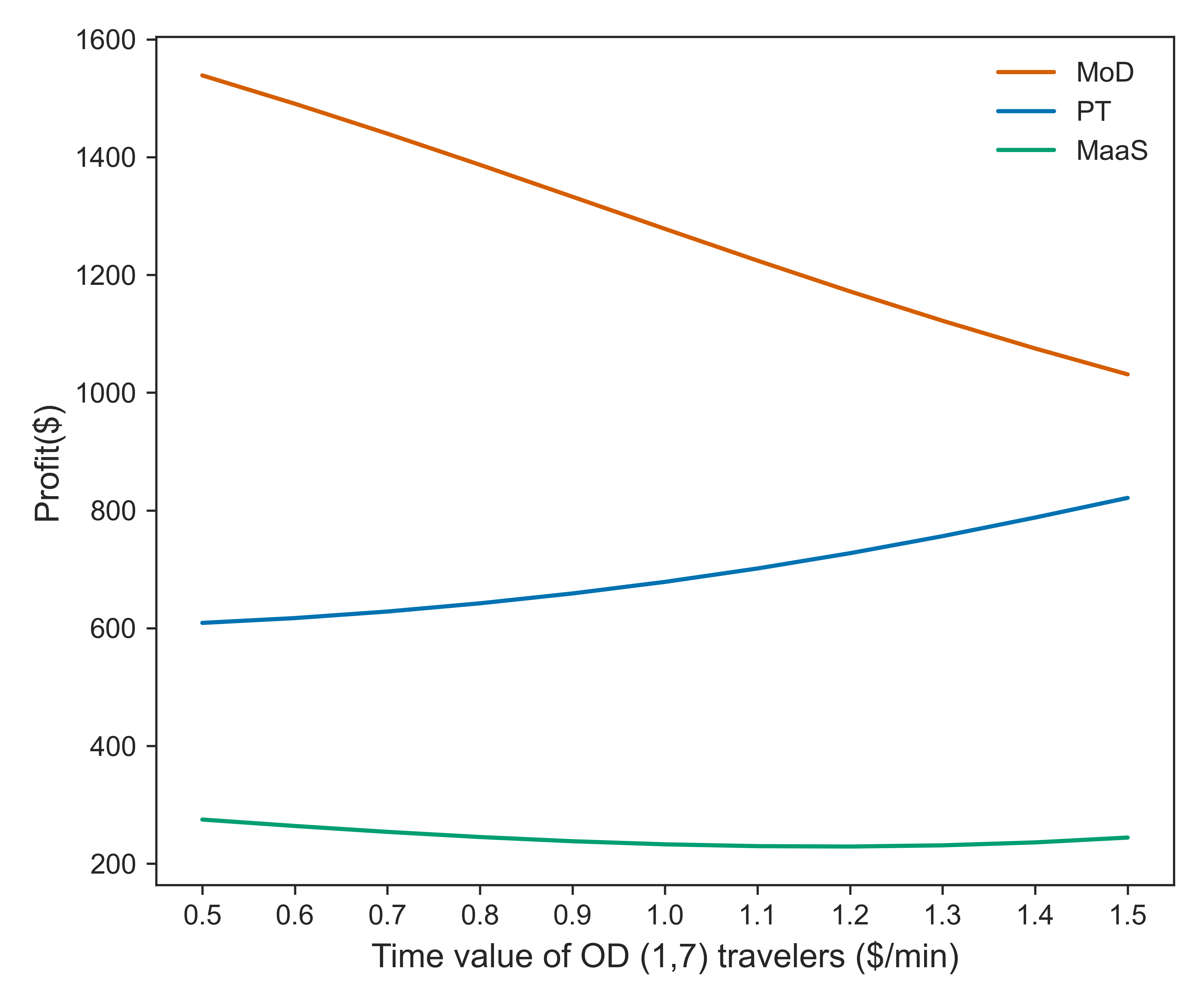}
    \caption{Operator's profit}
    \label{fig:VTT_sensitivity_profit}
  \end{subfigure}
  \caption{Market outcomes against the time value of OD pair (1,7).}
  \label{fig:VTT change}
\end{figure}

As an illustrative example, we vary $v_{17}$, the value of time for OD pair (1,7), and report the resulting equilibrium outcomes in Figure~\ref{fig:VTT change}. 
As shown in Figure~\ref{fig:VTT_sensitivity_split}, as $v_{17}$ increases, the proportion of travelers choosing private driving increases monotonically, which consequently reduces MaaS and non-MaaS demands between the same OD pair. 

To explore the network-wise modal shift, we plot the mode-specific total link flow in Figure~\ref{fig:VTT_sensitivity_flow}.
As $v_{17}$ increases, the total driving flow in the network rises, accompanied by a reduction in MoD flow and a moderate increase in PT flow. 
Notably, the increase in PT usage does not originate from travelers between OD (1,7), whose non-driving share declines as shown in Figure~\ref{fig:VTT_sensitivity_split}. 
Instead, it reflects network-wide adjustments induced by congestion effects. Such an adjustment is clearly illustrated in Figure~\ref{fig:VTT_sensitivity_TTPT}, which plots the travel time of road link (1,2) and the flow of PT link (1,2) of OD pair (1,5). 
The increased driving demand associated with OD (1,7) leads to higher congestion on road link (1,2), as reflected by the rising travel time. As a consequence, more travelers between OD pair (1,5), who also travel through link (1,2), shift toward PT and thus lead to the flow increase observed in Figure~\ref{fig:VTT_sensitivity_TTPT}. 
Finally, Figure~\ref{fig:VTT_sensitivity_profit} reports the impact of time value variations on operator profits. As expected, the reduction in MoD travel flows illustrated in Figure~\ref{fig:VTT_sensitivity_flow} leads to a substantial decrease in the MoD profit. 
In contrast, the PT operator experiences a moderate profit increase, driven by the additional PT demand. The MaaS platform’s profit is comparatively less affected, reflecting the fact that demand reductions and increases occur across different OD pairs and partially offset each other at the aggregate level.

Overall, this experiment illustrates how heterogeneity in travelers’ value of time can generate indirect and non-local effects through congestion interactions in the network. It suggests that a change in preferences for a single OD pair can propagate through shared network links, and consequently, induce modal shifts of travelers between other OD pairs and affect the profit of all service operators. Nevertheless, the MaaS platform shows more robustness towards the variation as it can retain the service quality by reallocating travel flows across the multi-modal network. 

}

{
\subsubsection{Sensitivity analysis on link capacity}


\begin{figure}[htb]
  \centering
  \begin{subfigure}[t]{0.48\textwidth}
    \centering
    \includegraphics[width=\linewidth]{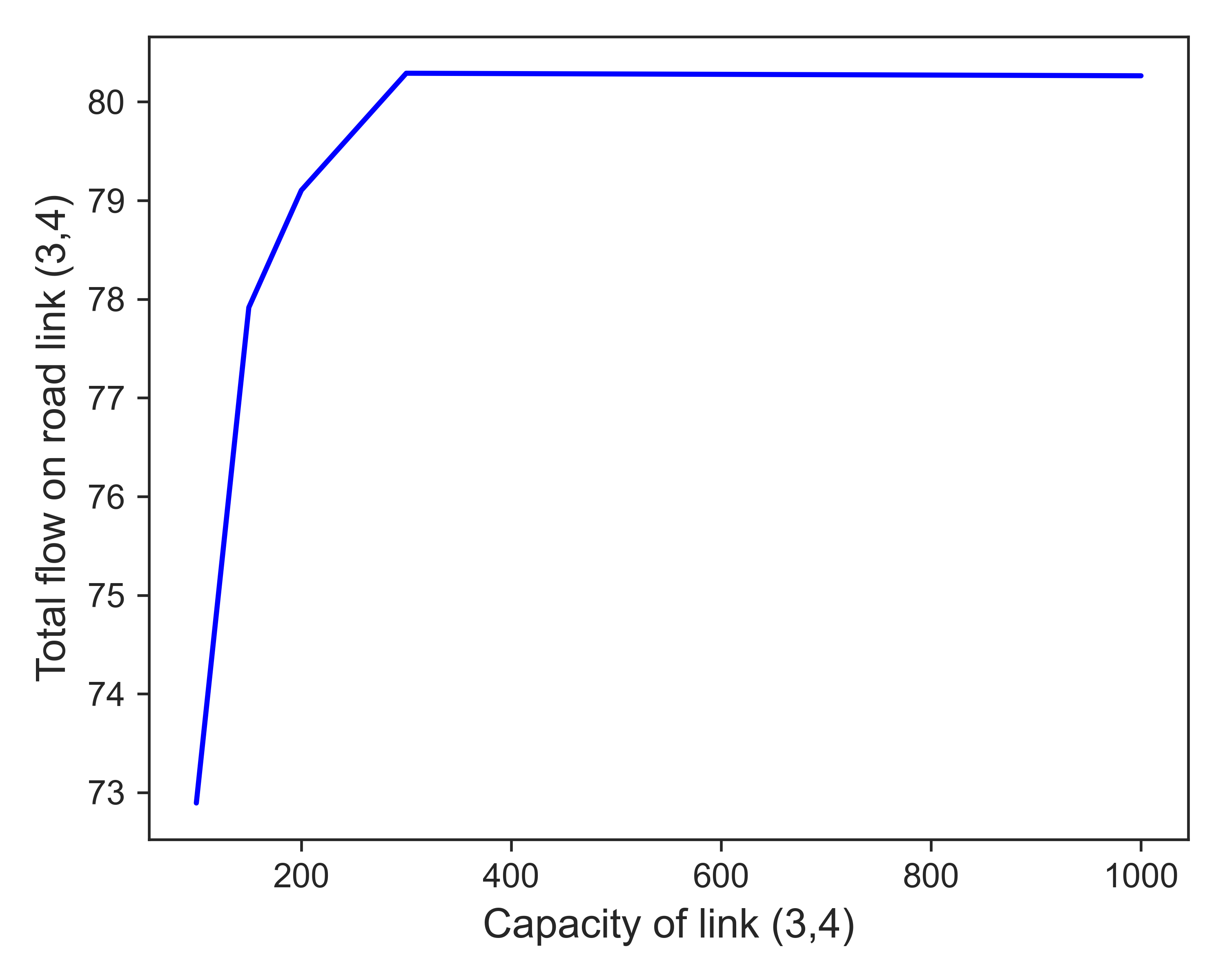}
    \caption{{Combined driving and MoD flows on link 3-4}}
    \label{fig:Cap_flow}
  \end{subfigure}
    \hfill
    \begin{subfigure}[t]{0.48\textwidth}
    \centering
    \includegraphics[width=\linewidth]{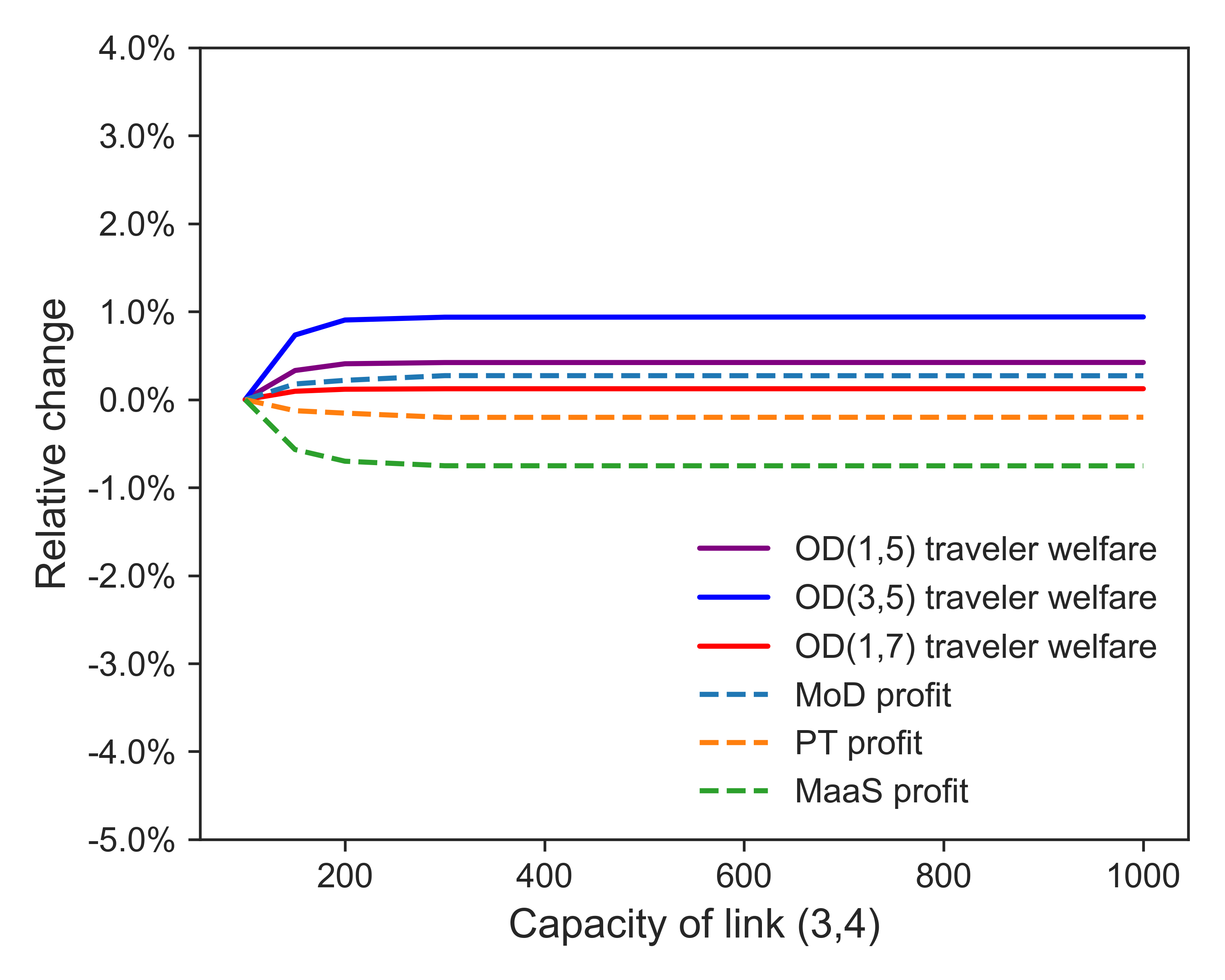}
    \caption{Operator profit and traveler welfare}
    \label{fig:Cap_welfare_profit}
    \end{subfigure}
    \hfill
    \begin{subfigure}[t]{0.48\textwidth}
    \centering
    \includegraphics[width=\linewidth]{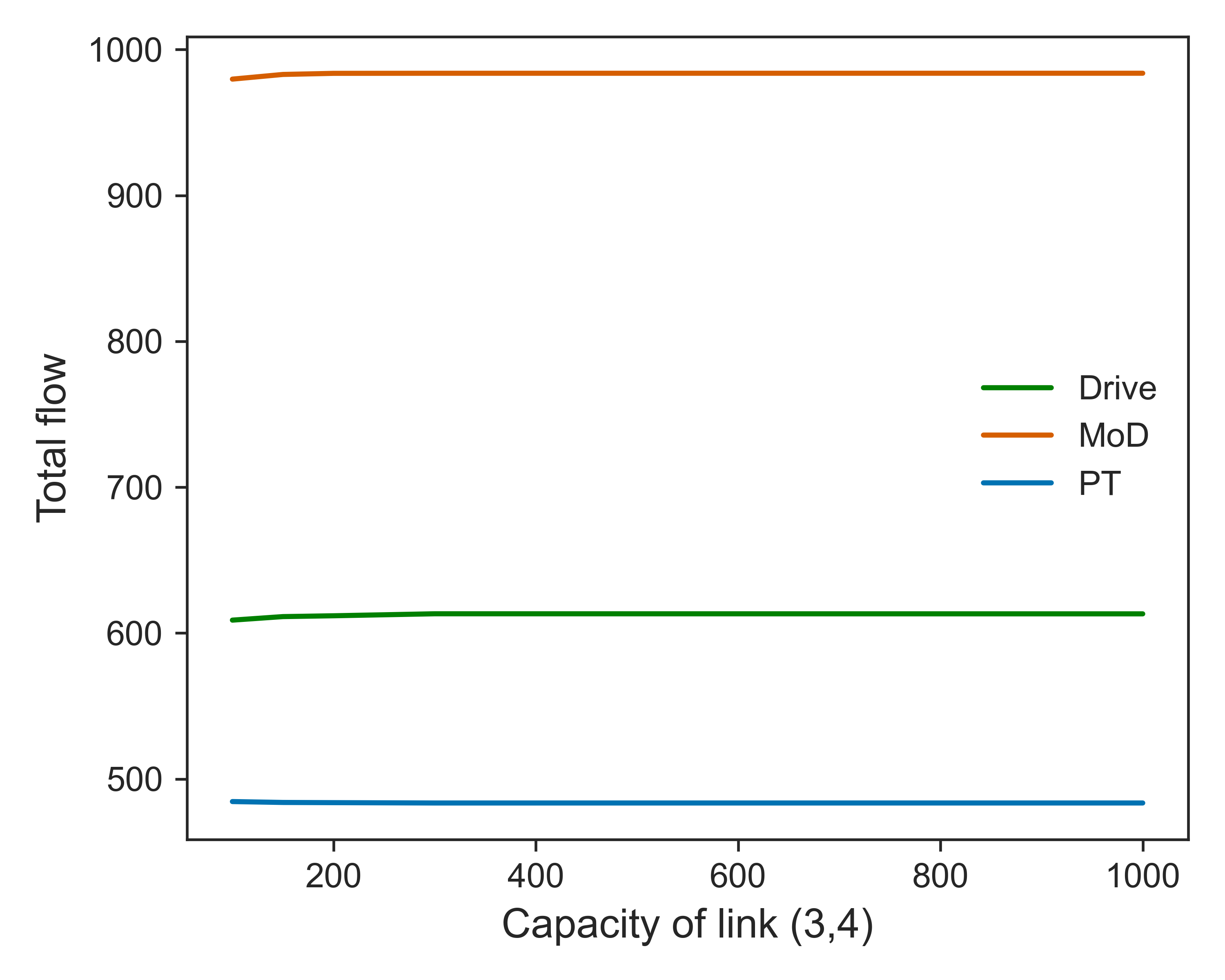}
    \caption{Mode-specific total link flows}
    \label{fig:Cap_total_flow}
  \end{subfigure}
  \caption{Market outcomes against the capacity of road link (3,4).}
  \label{fig:Cap change}
\end{figure}


This section proceeds to analyze the impact of link capacity on equilibrium outcomes. Similarly, we select road link (3,4), which is centrally located in the small network, and vary its capacity, while keeping all other parameters fixed. The resulting equilibrium outcomes are reported in Figure~\ref{fig:Cap change}.

As shown in Figure~\ref{fig:Cap_flow}, increasing the capacity of road link (3,4) initially leads to an increase in the total flow on this link (both driving and MoD). However, once the capacity exceeds approximately 350, the flow increase diminishes and eventually saturates. 
The corresponding impacts on traveler welfare and operator profits are reported in Figure~\ref{fig:Cap_welfare_profit}. 
Traveler welfare improves for all OD pairs as the link capacity increases, though the magnitude of improvement is modest and exhibits clear diminishing returns. 
In particular, the welfare gains flatten once the capacity exceeds approximately the same threshold at which link flows saturate. 
Changes in operator profits are also comparatively small: the MoD operator experiences a slight profit increase, whereas the PT operator and the MaaS platform incur marginal profit reductions, all within a narrow range of 1\%.
To further understand these limited impacts, we plot the total link flows by mode in Figure~\ref{fig:Cap_total_flow}. As shown, the total driving, MoD, and PT flows remain nearly unchanged across the tested capacity range, suggesting that the capacity expansion primarily redistributes traffic locally on link (3,4).

In sum, this experiment indicates that, at least in the tested network, capacity expansion of a single central link yields limited system-wide influences: the overall modal split remains largely unchanged, resulting in modest changes in traveler welfare and operator profits. 

}

\subsection{Sioux Falls network}\label{sec:experiment-large}

\begin{figure}[htb]
  \centering
  \includegraphics[width=0.8\textwidth]{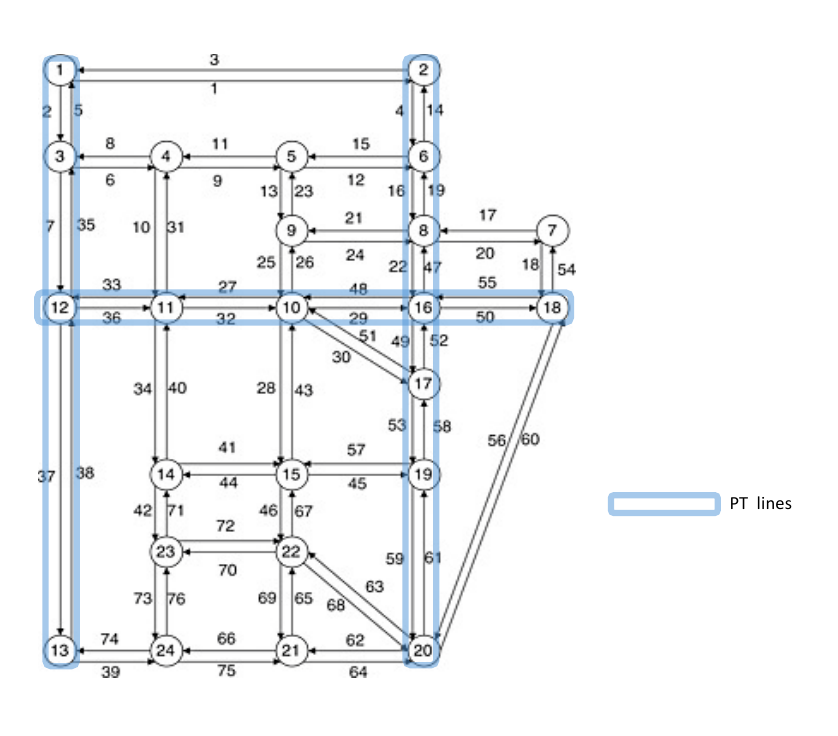} 
  \caption{Extended Sioux Falls network.}
  \label{fig:Sioux Falls network}
\end{figure}

To understand the impacts of MaaS in real-world networks, we augment the Sioux Falls network with three PT lines shown in Figure~\ref{fig:Sioux Falls network}, and construct the multi-modal transport network accordingly. 
The link travel time and price are specified in the same way as Section~\ref{sec:experiment-small-setup}, while the link parameters and exogenous variables are summarized in the Appendix.~\ref{sec:appendix- setting}

Following Section~\ref{sec:experiment-small-transfer}, we compare the system performance with and without the MaaS platform, assuming the MaaS platform leads to a major reduction of transfer penalty (see Appendix~\ref{sec:appendix- setting} for details). 
The aggregate performance metrics are reported in Table~\ref{tab:SF Comparison of results with and without MaaS}.
In line with the findings in the small network, the introduction of MaaS helps reduce private driving, and the drop is more significant in the Sioux Falls network, i.e., 8.2\% compared to 1.2\%. 
On the other hand, the MaaS platform wins a similar market share of 39.5\% and distributes a majority of passenger flows to PT, leading to a total of 18\% increase in PT demand. 
This significant increase is partially attributed to the surge in multi-modal travel, which is indicated by the considerable amount of transfer flow. 
On the other hand, the total share of MoD decreases substantially from 40\% to 31\% after MaaS enters the market. 
These findings again confirm the great potential of MaaS for replacing private driving and promoting sustainable travel modes.



Similar to the small network, the equilibrium on the Sioux Falls network reaches a ``win-win-win'' situation, where all service operators generate more profits, the MaaS platform is economically viable, and travelers enjoy a higher travel utility. Differently, the improvement in traveler welfare is far more promising in the Sioux Falls network, i.e., 14\% compared to 1.16\%. It also contributes to a majority of the growth in social welfare. 
Additionally, the profit growth for service operators is also considerable. Specifically, the majority of PT revenue comes from the MaaS payment, while MoD has a more balanced revenue composition.

\begin{table}[H]
  \centering
  \caption{Aggregate performance of Sioux Falls network with and without MaaS.}
  \label{tab:SF Comparison of results with and without MaaS}
  \begin{tabular}{lccc}
    \toprule
    & \textbf{Without MaaS} & \multicolumn{2}{c}{\textbf{With MaaS}} \\
    \midrule
    \textbf{Market share (\%)} & & &\\
    \quad non-MaaS & 51.5 & 20.2 & \\
    \quad \quad -MoD & 40.9 & 15.6 & \\
    \quad \quad -PT & 10.6  & 4.6 &\\
    \quad MaaS     & --     & 39.5 & \\
    \quad \quad -MoD & --&15.5 & \\
    \quad \quad -PT  & -- & 24.0 &\\
    \quad Driving  & 48.5 & 40.3 &$\textcolor{red}{\downarrow\,8.2\%}$ \\
    \textbf{Transfer flow}  & 0.87 & 189.14 &\\
    \midrule
    \textbf{Profit (\$)} & & \\
    \quad PT       & 337,286 & 375,636 & $\textcolor{blue}{\uparrow\,11.4\%}$ \\
    \quad MoD      & 221,756 & 376,355 & $\textcolor{blue}{\uparrow\,69.7\%}$ \\
    \quad MaaS     & --      & 5,658 & \\
    \textbf{Total Profit} & 559,043 & 757,649 & $\textcolor{blue}{\uparrow\,35.5\%}$ \\
    \midrule
    \textbf{non-MaaS price} & & &\\
    \quad PT     &3.65  & 2.42 & $\textcolor{red}{\downarrow\,33.7\%}$ \\
    \quad MoD    &0.81 & 1.04 & $\textcolor{blue}{\uparrow\,28.4\%}$\\
    \midrule
    \textbf{Traveler Welfare (\$)} & -9,934,124 & -8,545,142 & $\textcolor{blue}{\uparrow\,14.0\%}$\\
    \textbf{Social Welfare (\$)} & -9,375,081 & -7,787,493 & $\textcolor{blue}{\uparrow\,16.9\%}$ \\
    \bottomrule
  \end{tabular}
\end{table}




\section{Conclusions}\label{sec:conclusion}
In this paper, we consider a coopetitive MaaS system with different types of strategic players: the MaaS platform purchases service capacity from service operators and competes for travelers with an OD-based pricing scheme for profit; service operators use their remaining capacities to serve single-modal trips and optimize pricing strategies to maximize their profits; travelers make both mode choices and route choices in the multi-modal transportation network, subject to congestion and prices. 
We model this MaaS system as a multi-leader-multi-follower game that captures the complex strategic interactions among the MaaS platform, service operators, and travelers. 
Inspired by the dual formulation for traffic assignment problems, we further introduce a \textit{virtual} traffic operator, along with the MaaS platform and service operators. 
This enables us to simplify the bi-level multi-leader-multi-follower game into a single-level variational inequality (VI) problem. 
A key advantage of the VI formulation proposed in this study is that it supports parallel solution procedures: both the upper-level leaders' problems and lower-level followers' problems can be updated independently. Further, thanks to the PUMCM, the follower's best response corresponds to the optimal routing decisions in the multi-modal network and possesses a closed-form expression, which enables large-scale computations.
In addition, with the connection between VI formulation and equilibrium conditions, we are able to prove the existence of equilibrium under mild assumptions.

We demonstrate the modeling framework and solution algorithm using a toy network and an augmented Sioux Falls network. The small network also allows us to further explore the competition between the MaaS platform and service operators, as well as the key factors that influence the economic viability and the condition of Pareto-improving. 
Our numerical results suggest that MaaS has the potential to substitute part of the driving trips and promote multi-modal travels. In addition, MaaS can simultaneously improve the profits of service operators and the welfare of travelers, meanwhile sustaining its own business. Consequently, the introduction of MaaS creates a ``win-win-win'' situation. Our sensitivity analysis further reveals that such a Pareto-improving situation, however, depends on the negotiated wholesale capacity price. 
While a low wholesale price favors the MaaS platform and enhances traveler welfare, it undermines the profitability of service operators. It thus exists a Pareto-improving regime of the wholesale price, within which all stakeholders could benefit from the introduction of MaaS. 
We further demonstrate these main findings with the Sioux Falls network, and validate the scalability of the proposed solution algorithm. 

There are several potential directions to extend from here. First, future research can explore conditions for the existence of a Pareto-improving wholesale price regime and design mechanisms that induce the Pareto-improving regime in line with system-level objectives (e.g., maximizing total operator profits). 
Besides, the current model assumes the total demand is fixed. The assumptions no longer hold if we consider the continuously evolving travel demand pattern and new emerging mobility services. A natural future direction is to incorporate elastic and heterogeneous demands into the model.
{Moreover, since the proposed model generally yields multiple equilibria, future work could focus on the equilibrium selection mechanism that characterizes and compares different equilibrium outcomes with additional intervention to induce a unique and desirable equilibrium.}
Finally, as our model currently involves only two types of operators and a single MaaS platform, extending it to include multiple MaaS platforms with diverse objectives and operational strategies would enhance its applicability to more realistic multi-player mobility systems. 

\section*{Acknowledgments}
The work was funded by the Swiss National Science Foundation (219232).

\bibliographystyle{apalike} 
\bibliography{reference}
\appendix

\section{Detailed experiment setup}\label{sec:appendix- setting}
For the experiments on small network in Section~\ref{sec:experiment-small-setup}, the values of regular link parameters are reported in Table~\ref{tab:physical_links}.

\begin{table}[H]
\centering
\caption{Physical Links in Small Networks.}
\label{tab:physical_links}
\renewcommand{\arraystretch}{1.2}
\begin{tabular}{c|c|c|c|c}
\toprule
\textbf{Link} & $L_a$ & $K_a$ & \textbf{PT: $T_a$} & \textbf{MoD: $T_a$} \\
\midrule
(1, 2)  & 2.0 & 300 & 7.5 & 5.0 \\
(2, 3)  & 1.6 & 200 & 6.0 & 4.0 \\
(2, 7)  & 4.0 & 300 & --  & 10.0 \\
(3, 4)  & 2.0 & 300 & 7.5 & 5.0 \\
(3, 5)  & 2.0 & 300 & 7.5 & 5.0 \\
(3, 6)  & 3.2 & 200 & --  & 8.0 \\
(4, 5)  & 2.4 & 300 & --  & 6.0 \\
(4, 7)  & 2.0 & 300 & 7.5 & 5.0 \\
(5, 6)  & 2.0 & 300 & 7.5 & 5.0 \\
(6, 4)  & 2.4 & 400 & --  & 6.0 \\
(6, 8)  & 2.0 & 300 & --  & 5.0 \\
(7, 8)  & 2.0 & 300 & --  & 5.0 \\
\bottomrule
\end{tabular}
\end{table}

For the experiment on Sioux Falls network in Section~\ref{sec:experiment-large}, the values of regular link parameters are reported in Table~\ref{tab:SiouxFallsNetwork}. And all links are bidirectional. The OD pairs of demands are set to be 1.2 times Sioux Falls benchmark~\citep{bar2016transportation}.
We apply a free-flow time factor of 1.2 to public transit (PT) services, indicating a longer base travel time compared to regular links. 
\begin{longtable}{c|c|c|c|c}
\caption{Link Attributes of Sioux Falls network.} \label{tab:SiouxFallsNetwork} \\
\toprule
\textbf{Link} & \textbf{MoD $K_a$} & \textbf{PT $K_a$} & $T_a$ & \textbf{$L_a$} \\
\midrule
\endfirsthead

\multicolumn{5}{l}{\small\slshape Continued from previous page} \\
\toprule
\textbf{Link} & \textbf{MoD $K_a$} & \textbf{PT $K_a$} & $T_a$ & \textbf{$L_a$} \\
\midrule
\endhead

\midrule
\multicolumn{5}{r}{\small\slshape Continued on next page} \\
\endfoot

\bottomrule
\endlastfoot

(1,2)   & 25900.2  & --         & 6 & 1.2 \\
(1,3)   & 23403.5  & 1,000,000  & 4 & 0.8 \\
(2,6)   & 4958.18  & 1,000,000  & 5 & 1.0 \\
(3,4)   & 17110.5  & --         & 4 & 0.8 \\
(3,12)  & 23403.5  & 1,000,000  & 4 & 0.8 \\
(4,5)   & 17782.8  & --         & 2 & 0.4 \\
(4,11)  & 4908.83  & --         & 6 & 1.2 \\
(5,6)   & 4948     & --         & 4 & 0.8 \\
(5,9)   & 10000    & --         & 5 & 1.0 \\
(6,8)   & 4898.59  & 1,000,000  & 2 & 0.4 \\
(7,8)   & 7841.81  & --         & 3 & 0.6 \\
(7,18)  & 23403.5  & --         & 2 & 0.4 \\
(8,9)   & 5050.19  & --         & 10 & 2.0 \\
(8,16)  & 5045.82  & 1,000,000  & 5 & 1.0 \\
(9,10)  & 13915.8  & --         & 3 & 0.6 \\
(10,11) & 10000    & 1,000,000  & 5 & 1.0 \\
(10,15) & 13512    & --         & 6 & 1.2 \\
(10,16) & 4854.92  & 1,000,000  & 4 & 0.8 \\
(10,17) & 4993.51  & --         & 8 & 1.6 \\
(11,12) & 4908.83  & 1,000,000  & 6 & 1.2 \\
(11,14) & 4876.51  & --         & 4 & 0.8 \\
(12,13) & 25900.2  & 1,000,000  & 3 & 0.6 \\
(13,24) & 5091.26  & --         & 4 & 0.8 \\
(14,15) & 5127.53  & --         & 5 & 0.8 \\
(14,23) & 4924.79  & --         & 4 & 0.8 \\
(15,19) & 14564.8  & --         & 3 & 0.6 \\
(15,22) & 9599.18  & --         & 3 & 0.6 \\
(16,17) & 5229.91  & 1,000,000  & 2 & 0.4 \\
(16,18) & 19679.9  & 1,000,000  & 3 & 0.6 \\
(17,19) & 4823.95  & 1,000,000  & 2 & 0.4 \\
(18,20) & 23403.5  & --         & 4 & 0.8 \\
(19,20) & 5002.61  & 1,000,000  & 4 & 0.8 \\
(20,21) & 5059.91  & --         & 6 & 1.2 \\
(20,22) & 5075.7   & --         & 5 & 1.0 \\
(21,22) & 5229.91  & --         & 2 & 0.4 \\
(21,24) & 4885.36  & --         & 3 & 0.6 \\
(22,23) & 5000     & --         & 4 & 0.8 \\
(23,24) & 5078.51  & --         & 2 & 0.4 \\
\end{longtable}

The following parameter settings are used in the experiments:
\begin{itemize}
    \item \textbf{Transfer and access link settings:}
        The free-flow time for MoD access links is set to 1, while for PT transfers is set to 2 in both Section~\ref{sec:experiment-small-transfer} and ~\ref{sec:experiment-large}.
        The link capacity is set to 100 for MoD access links and 500 for PT access links in Section~\ref{sec:experiment-small-transfer}. In the Sioux Falls network, the capacities are set to 60,000 for MoD access and 30,000 for PT access in Section~\ref{sec:experiment-large}.
    \item \textbf{Driving cost parameters:}
        In the small network, the fixed cost of driving is set to 5, and in the Sioux Falls network, it is set to 2.
        The variable cost per unit distance is set to 0.2 for both network experiments.
    \item \textbf{Transfer penalty:}
        The transfer penalty is set to 10 for non-MaaS, 1 for MaaS with major reduction, and 3 for MaaS with moderate reduction in Section~\ref{sec:experiment-small-transfer}.
        The transfer penalty for non-MaaS and MaaS scenario is respectively 8 and 0.5 in Section~\ref{sec:experiment-large}. 
    \item \textbf{Wholesale pricing:}
        The wholesale capacity price is fixed at 1.3 for both PT and MoD in Section~\ref{sec:experiment-small-transfer}, and 1.0 for PT and 1.5 for MoD in Section~\ref{sec:experiment-large}
    \item \textbf{BPR function parameters:}
        The Bureau of Public Roads (BPR) function is applied with parameters $\alpha = 0.15$ and $\beta = 4$.
\end{itemize}


The parameters in Algorithm for experiments are set as follows:
\begin{itemize}
    \item \textbf{Step size:} $\alpha =1\times 10^{-4}$;
    \item \textbf{Stopping threshold:} $\epsilon =1\times 10^{-6}$;
    \item \textbf{Residual function:} $R(y)=\|y-P_\Omega(y-\alpha F(y))\|$.
\end{itemize}

The algorithm is implemented in Python and with analytical gradient and best response sensitivity.




\end{document}